\documentclass[lettersize,journal]{IEEEtran}

\usepackage[caption=false,font=normalsize,labelfont=sf,textfont=sf]{subfig}
\usepackage[style=ieee,maxbibnames=99,sorting=none,giveninits=true,date=year,isbn=false,backend=bibtex]{biblatex}
\renewbibmacro{in:}{\ifentrytype{article}{}{\printtext{\bibstring{in}\intitlepunct}}}%对于article类不显示"In:"

\usepackage{array}%需要用到该宏包
\usepackage{footnote}
\makesavenoteenv{tabular}
\usepackage{caption}
\usepackage{graphicx}
\usepackage{float}
\usepackage{amsthm,amsmath,amssymb,bm}
\usepackage{mathrsfs}
\usepackage{xcolor}
\usepackage{dutchcal}
\usepackage{pifont}
\usepackage{makecell}
\usepackage{wasysym}
\usepackage{multirow} % 导入multirow宏包

\begin{document}

\title{Smartphone User Fingerprinting on Wireless Traffic}

\author{Yong~Huang,~\IEEEmembership{Member,~IEEE,}
        Zhibo~Dong,
        Xiaoguang~Yang, % <-this % stops a space
        Dalong~Zhang, 
        Qingxian~Wang,
        and
        Zhihua~Wang,~\IEEEmembership{Member,~IEEE} 
% <-this % stops a space
\thanks{Part of this work has been presented at ICIC 2024~\cite{yxg}.}

\thanks{This work was supported in part by the National Natural Science Foundation of China with Grant 62301499 and U22A2001, and the Henan Association for Science and Technology with Grant 2025HYTP037 \textit{(Corresponding author: Dalong Zhang.)}}

\thanks{Y. Huang, Z. Dong, X. Yang, D. Zhang, Q. Wang, and Z. Wang are with the School of Cyber Science and Engineering, Zhengzhou University, Zhengzhou 450001, China (e-mail: yonghuang@zzu.edu.cn; dongzhibo@gs.zzu.edu.cn; xiaoguangyang@gs.zzu.edu.cn; iedlzhang@zzu.edu.cn; wqx196008@163.com; zhwang@zzu.edu.cn).}
}

\maketitle

\begin{abstract}
Due to the openness of the wireless medium, smartphone users are susceptible to user privacy attacks, where user privacy information is inferred from encrypted Wi-Fi wireless traffic.
Existing attacks are limited to recognizing mobile apps and their actions and cannot infer the smartphone user identity, a fundamental part of user privacy.
To overcome this limitation, we propose U-Print, a novel attack system that can passively recognize smartphone apps, actions, and users from over-the-air MAC-layer frames.
We observe that smartphone users usually prefer different add-on apps and in-app actions, yielding different changing patterns in Wi-Fi traffic. 
U-Print first extracts multi-level traffic features and exploits customized temporal convolutional networks to recognize smartphone apps and actions, thus producing users' behavior sequences. 
Then, it leverages the silhouette coefficient method to determine the number of users and applies the k-means clustering to profile and identify smartphone users. 
We implement U-Print using a laptop with a Kali dual-band wireless network card and evaluate it in three real-world environments.
U-Print achieves an overall accuracy of 98.4\% and an F1 score of 0.983 for user inference.
Moreover, it can correctly recognize up to 96\% of apps and actions in the closed world and more than 86\% in the open world.
\end{abstract}


% \fi
\IEEEpeerreviewmaketitle

\section{Introduction}
\IEEEPARstart{I}{n} the recent decade, smartphones have become an integral part of people's everyday lives. 
It is reported that in 2023, smartphone users averagely check their smartphones 144 times and spend 4 hours and 25 minutes on smartphones daily in the United States~\cite{phoneusage}.
By providing seamless connectivity to others, smartphones have reshaped our society in many aspects, including work, entertainment, social communication, and so on~\cite{zhu2013mobile}.
Nowadays, smartphone users prefer the Internet connection via a Wi-Fi network wherever it is available, such as homes, offices, and hotels, for cheap and fast connectivity. 
Although Wi-Fi networks generally adopt encryption standards like WPA2 and WPA3 to secure data confidentiality, they are still susceptible to user privacy attacks that infer user privacy from encrypted wireless traffic~\cite{pang200711,wang2015know,zhai2021identify}.
Since smartphones carry a considerable amount of personal privacy information~\cite{stober2013you,taylor2017robust,tu2018your,conti2015can,li2017demographic,7293229}, such as gender, age, and hobbies~\cite{aceto2019mobile,aceto2019mimetic,chen2017powerful}, it is of great importance to investigate the extent to which smartphones are vulnerable to such attacks~\cite{bozorgi2022still,bakopoulou2021fedpacket,li2023meta}.

Although extensive endeavors have been devoted to user privacy attacks on smartphones, existing approaches still fall short in inferring smartphone user identities, an essential part of user privacy.  
As shown in Table~\ref{comparing with others}, AppScanner~\cite{taylor2016appscanner}, FLOWPRINT~\cite{van2020flowprint}, and FOAP~\cite{li2022foap} classified apps or in-app actions performed by smartphone users from the side channel of TCP/IP layer traffic. 
However, the above techniques required direct connection to access points, routers, or switches to obtain TCP/IP layer traffic, which is difficult to achieve due to the existence of physical isolation and other protective measures. 
Hence, the recent work~\cite{atkinson2018your,wang2015know,zhang2011inferring,li2022packet} focused on attacking MAC layer traffic that can be obtained by sniffing over-the-air 802.11 frames.
Several approaches~\cite{atkinson2018your,wang2015know,zhang2011inferring} realized app classification under the closed-world assumption, where apps must be presented in both the training and testing phases. 
However, there are millions of apps in mobile app stores~\cite{appnum}, rendering it impossible to build fingerprints for all of them.
While PACKETPRINT~\cite{li2022packet} recognized apps and actions in the open world, it cannot infer the identities of smartphone users.
Hence, none of the existing approaches simultaneously meet the requirements of recognizing smartphone apps, actions, and users on Wi-Fi MAC layer traffic in the open world.

\begin{table}[]
\caption{Comparison with Other User Privacy Attacks.}
\begin{tabular}{cccccc}
\hline
\textbf{\begin{tabular}[c]{@{}c@{}}User Privacy\\ Attacks\end{tabular}}  & \textbf{\begin{tabular}[c]{@{}c@{}}MAC \\ Layer\end{tabular}} & \textbf{\begin{tabular}[c]{@{}c@{}}App\end{tabular}} & \textbf{\begin{tabular}[c]{@{}c@{}}Action\end{tabular}} & \textbf{\begin{tabular}[c]{@{}c@{}}User\end{tabular}} & \textbf{\begin{tabular}[c]{@{}c@{}}Open\\ World\end{tabular}}\\
\hline
\begin{tabular}[c]{@{}c@{}}AppScanner~\cite{taylor2016appscanner}\end{tabular}         &  \ding{56} & \ding{52} & \ding{56} & \ding{56} & \ding{56}\\
\hline
\begin{tabular}[c]{@{}c@{}}FLOWPRINT~\cite{van2020flowprint}\end{tabular} & \ding{56} & \ding{52} & \ding{56} & \ding{56} & \ding{56}\\
\hline
\begin{tabular}[c]{@{}c@{}}FOAP~\cite{li2022foap}\end{tabular} & \ding{56} & \ding{52} & \ding{52} & \ding{56} & \ding{52}\\                                          
\hline
\begin{tabular}[c]{@{}c@{}}PACKETPRINT~\cite{li2022packet}\end{tabular}  &  \ding{52} & \ding{52} & \ding{52} & \ding{56} & \ding{52}\\
\hline
\begin{tabular}[c]{@{}c@{}}Wang et al.~\cite{wang2015know}\end{tabular}  &  \ding{52} & \ding{52} & \ding{56} & \ding{56} & \ding{56}\\
\hline
\begin{tabular}[c]{@{}c@{}}Zhang et al.~\cite{zhang2011inferring}\end{tabular} & \ding{52} & \ding{52} & \ding{56} & \ding{56} & \ding{56}\\
\hline
\begin{tabular}[c]{@{}c@{}}John et al.~\cite{atkinson2018your}\end{tabular} & \ding{52} & \ding{52} & \ding{56} & \ding{56} & \ding{56}\\
\hline
\begin{tabular}[c]{@{}c@{}}\textbf{U-Print (Ours)}\end{tabular} & \ding{52} & \ding{52} & \ding{52} & \ding{52} & \ding{52}\\
\hline
\label{comparing with others}
\end{tabular}
\end{table}
This paper presents U-Print, a novel attack system that profiles smartphone users via encrypted Wi-Fi MAC frames.
We observe that smartphone users are tempted to install different add-on apps, such as YouTube and Instagram.
Moreover, even in the same app, users prefer different in-app actions, such as text chat and voice chat in WhatsApp. 
Due to the differences in app communication protocols and specifications, users' app and action preferences would trigger different changing patterns in Wi-Fi traffic.
Based on this observation, we can build profiles for smartphone users by inferring their preferred mobile apps and in-app actions and then identify them given new wireless traces.
It is worth noting that our objective is to demonstrate how smartphone users' usage habits or preferences are exposed in daily used Wi-Fi networks, and further to raise public awareness of such risks and lay the foundation for future research on protection mechanisms.

We realize the above idea by addressing the following three challenges.

1) \textit{How to extract fine-grained features from encrypted wireless traffic?} 
In a Wi-Fi network, application-layer messages are encrypted in the payload part of an 802.11 data frame, leaving little information about apps and actions performed on users' smartphones. 
To deal with this issue, we first segment raw wireless traffic into effective traffic traces and represent them using MAC-layer plaintext metadata, including frame time, size, and direction.
Then, various statistical and contextual features are extracted with different sliding windows to generate app and in-app action feature samples.

2) \textit{How to perform accurate app and action classification in the open world?}
Most of the existing approaches~\cite{baek2023targeted,li2022foap,ni2023eavesdropping,aceto2020toward} rely on manual annotation for traffic traces and adopt the closed-world assumption~\cite{aceto2018multi,wang2020automatic}, which yields poor classification performance in real-world environments.
To avoid this problem, we leverage temporal convolutional networks (TCNs) with OpenMax functions to facilitate app and action classification in the open world.
Moreover, an Android app called the label recorder is developed to run on smartphones to facilitate fine-grained sample annotation for classifier training.

3) \textit{How to achieve effective user profiling under MAC randomization?} 
Nowadays, most smartphones adopt the MAC randomization strategy when connecting to Wi-Fi networks~\cite{macrandom}.
In this condition, one smartphone will have multiple randomized MAC addresses throughout the profiling process, rendering MAC addresses unlinkable to user identities.
To deal with this challenge, we extract unique behavior samples that reflect the users' usage habits from the app and action predictions.
Then, the silhouette coefficient method (SCM) is applied to determine the number of victim users in the targeted area.
Finally, we use the k-means clustering algorithm to build user profiles in the profiling phase and identify them in the inference phase.  

We implement U-Print using a Lenovo laptop connecting to a Kali dual-band wireless network card and evaluate it with 40 mobile apps and 12 smartphone users in three different environments.
The evaluation results demonstrate that U-Print achieves an overall accuracy of 98.4\% and an F1 score of 0.983 for user inference, requiring only about 16 seconds for each inference. 
Moreover, it can correctly classify up to 96\% of apps and actions in the closed world and more than 86\% in the open world.

The main contributions are summarized as follows.
\begin{itemize}
    \item We demonstrate that user preferences on mobile apps and in-app actions can be reflected in Wi-Fi traffic, thus providing an exciting opportunity for smartphone user fingerprinting.
    \item We propose U-Print that can passively infer users' add-on apps, in-app actions as well as identities in the open world from encrypted Wi-Fi traffic.
    \item We implement U-Print and conduct extensive experiments with 40 mobile apps and 12 smartphone users in three different environments to demonstrate the effectiveness of our system.
\end{itemize}

\begin{figure}
    \centering
    \includegraphics[width=1.0\linewidth]{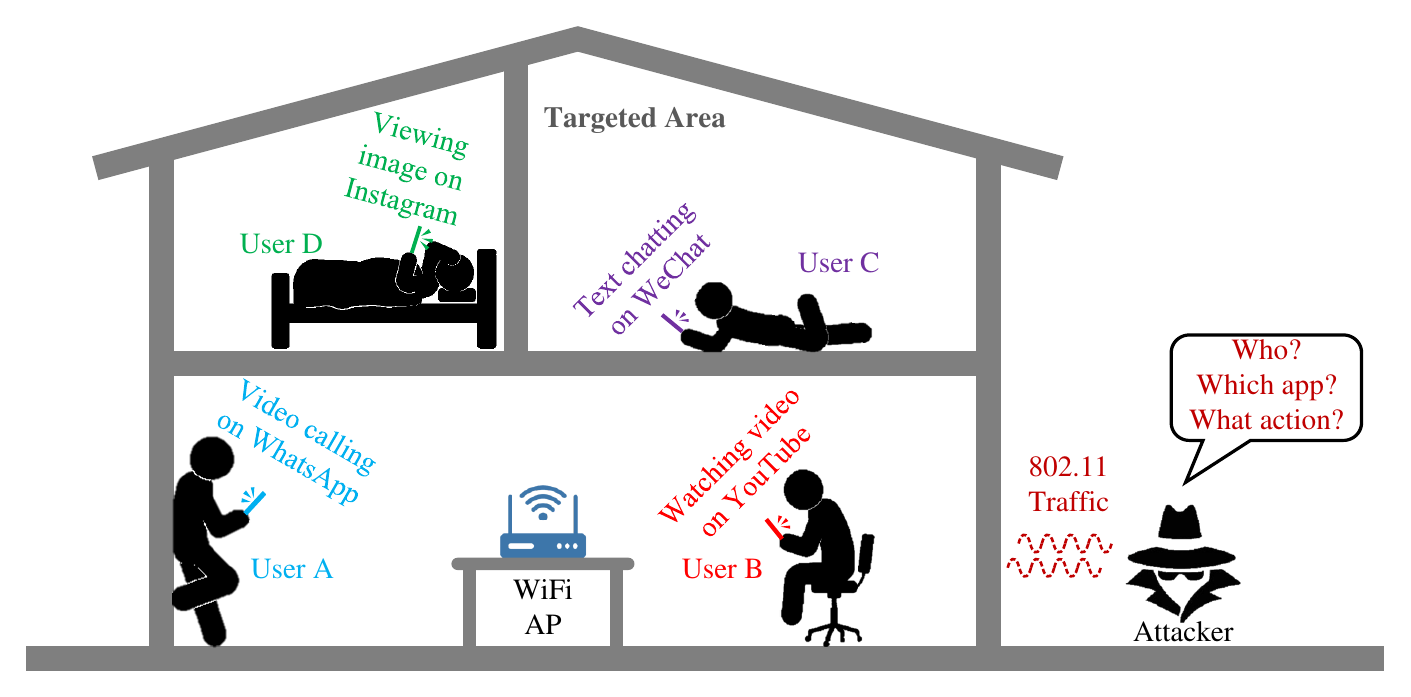}
    \caption{Threat model.}
    \label{fig:threatmodel}
\end{figure}

\section{Threat Model and Feasibility Study}

\subsection{Threat Model}
We consider a common scenario where a Wi-Fi network is installed in an indoor area, such as a home or an office, to provide Internet access as depicted in Fig.~\ref{fig:threatmodel}. 
In this network, one access point (AP) can be deployed for sufficient wireless connectivity within this area. 
Multiple users could exist in the area, such as many family members living in a house.
In such places, people attempt to exhibit routine smartphone usage patterns. 
For example, one user may prefer nightly entertainment at home and pervasive socializing at the office.

In this scenario, we consider user fingerprinting attacks by observing the network's wireless encrypted traffic. 
Specifically, the adversary deploys sniffing tools within the reception range of the Wi-Fi network but outside the targeted area in advance.
Since each AP would periodically transmit beacon frames with a consistent MAC address, the adversary can leverage Wireshark or Aircrack-ng to passively capture Wi-Fi MAC frames from and to the targeted AP over the air.
His goal is to build profiles for mobile users from collected wireless traffic and identify them given new traffic traces.
We assume that the adversary can access MAC-layer plaintext metadata information, such as destination address, source address, and arrival time in received IEEE 802.11 frames.
However, he does not have the network's password, and thus cannot obtain IP- or transport-layer headers protected by encryption protocols, such as WPA2 or WPA3~\cite{apthorpe2017closing}. 
Moreover, he cannot directly associate wireless traces with users based on MAC addresses due to the widely adopted MAC randomization scheme in smartphones~\cite{macrandom}.

\begin{figure}
    \centering
    \includegraphics[width=1\linewidth]{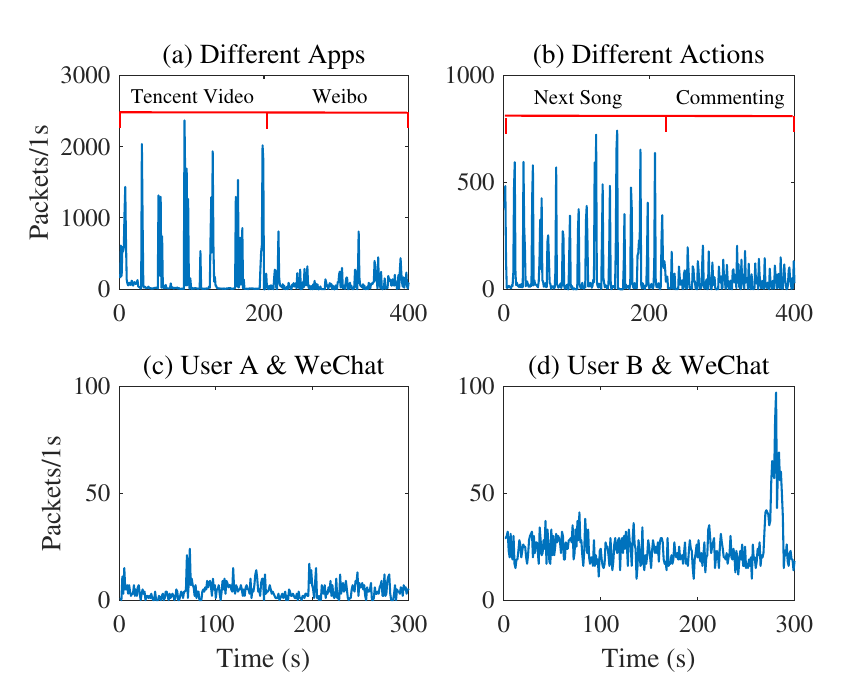}  
    \caption{Feasibility study.}
    \label{fig:FeasibilityStudy}
\end{figure}

\subsection{Feasibility Study}
We conduct some preliminary studies to verify the feasibility of identifying mobile users based on wireless encrypted traffic.  
To do this, we leverage a laptop with Wireshark to record 802.11 frames between a Xiaomi Mi~8 smartphone and a TP-Link WDR5620 Wi-Fi router.  
First, we run two apps, i.e., Tencent Video and Weibo, consecutively on the smartphone.  
Fig.~\ref{fig:FeasibilityStudy}~(a) shows the traffic rates of the smartphone in this setting. 
It can be observed that when Tencent Video is used, the traffic rate is fast and exhibits a clear periodicity in bursts.
However, when Weibo runs on a smartphone, the rate is relatively small and the bursts are compact.
The results indicate that each app has a unique traffic pattern.
Next, we run NetEase Cloud Music on the smartphone and conduct two actions, namely switching to the next song and commenting, respectively. 
As depicted in Fig.~\ref{fig:FeasibilityStudy}~(b), when the user is switching to the next song, the traffic bursts are high but relatively sparse. When it comes to commenting, the traffic rate becomes slow yet frequent, suggesting that different actions of the same app can incur different traffic changes.  
Based on the above results, we take a further step to verify the possibility of differentiating mobile users based on wireless traffic. 
To achieve this, two users, namely User A and User B, are successively required to use WeChat, an instant messaging app, on this smartphone in five minutes.
As shown in Fig.~\ref{fig:FeasibilityStudy}~(c), the traffic rate generated by User A is relatively low.
It may be due to his preference for text chat that incurs a low volume of application-layer messages.
However, as depicted in Fig.~\ref{fig:FeasibilityStudy}~(d), User B has a high average traffic rate, because he is more inclined to voice chat, generating lots of wireless frames.
Hence, even if two users use the same app, their usage habits could be different, resulting in different traffic patterns.  

\section{System Design}

\begin{figure*}
    \centering
    \includegraphics[width=0.95\linewidth]{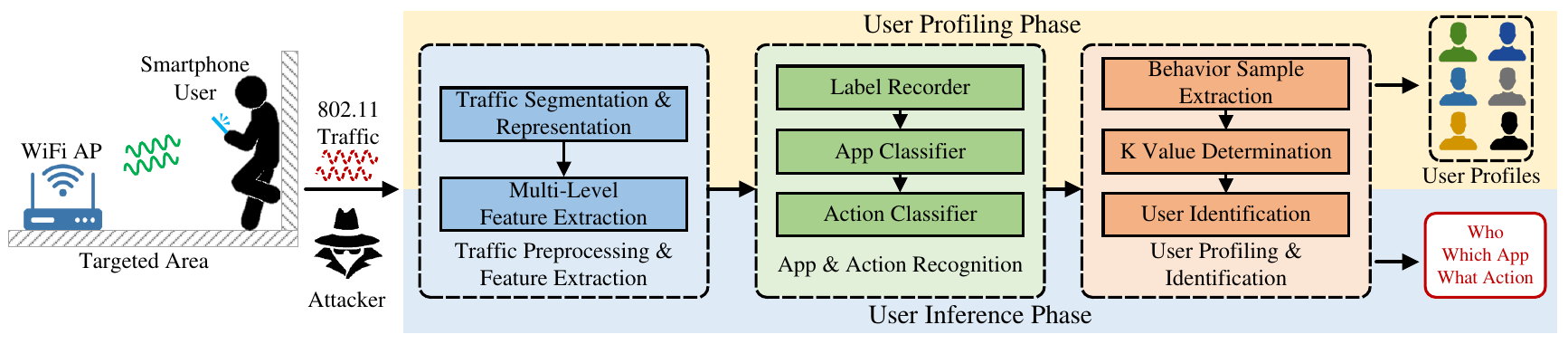}
    \caption{System overview.}
    \label{fig:SystemOverview}
\end{figure*}

\subsection{System Overview}
U-Print is a novel system that profiles mobile users via encrypted Wi-Fi MAC frames.
As depicted in Fig.~\ref{fig:SystemOverview}, U-Print runs on a computer with a wireless network card and can be deployed around targeted areas to sniff 802.11 traffic over the air.
U-Print undergoes two phases. 
In the profiling phase, it obtains a collection of wireless traffic generated by victim users and profiles users' app usage habits.
In the inference phase, it recognizes who is using which app with which action on her/his smartphone, given new sniffed wireless frames.
U-Print exploits the side channel of Wi-Fi communications to infer victims' privacy, making it completely passive and undetectable.
Furthermore, the generated user profiles can be leveraged to speculate user age, gender, health status, sexual orientation, and other private information~\cite{wang2015know}.
Note that our system can also use multiple sniffers or network interface cards to capture traffic from multiple APs or users within the targeted area and improve data fidelity.

As shown in Fig.~\ref{fig:SystemOverview}, U-Print consists of three core components -- \textit{Traffic Preprocessing and Feature Extraction}, \textit{App and Action Recognition}, and \textit{User Profiling and Identification}.
\begin{itemize}
    \item \textbf{Traffic Preprocessing and Feature Extraction.} First, this component segments raw wireless traffic into effective traces and represents them using useful metadata encapsulated in MAC frames. 
    Then, multi-level features are extracted with sliding windows for characterizing different apps and in-app actions. 
    \item \textbf{App and Action Recognition.} Based on the extracted feature samples, our system exploits temporal convolutional networks with OpenMax functions to recognize mobile apps and in-app actions in the open world. 
    Moreover, an Android app called the label recorder is developed to run on smartphones and facilitate fine-grained sample annotation for model training.
    \item \textbf{User Profiling and Identification.} In this component, our system converts app and action predictions into user behavior samples that reflect the unique usage patterns of different users.
    Then, the silhouette coefficient method is used to determine the number of smartphone users in the targeted area. 
    After that, our system exploits the k-means clustering for user profiling and identification.
\end{itemize}

\subsection{Traffic Preprocessing and Feature Extraction}
\textbf{Traffic Segmentation and Representation.} The first step of U-Print is to sniff Wi-Fi traffic from the targeted area and segment effective traffic traces that are generated by mobile users. 
To achieve this, U-Print sets the wireless network card to the monitoring mode to scan all Wi-Fi channels.
Once 802.11 frames are detected, it utilizes tools such as Wireshark and Aircrack-ng to capture them on the corresponding wireless channels. 
Due to apps running in the background, smartphones can also generate traffic without user interaction. 
It is necessary to segment out traffic traces incurred by users. 
Generally, the traffic rate of background apps is much smaller when compared with that of foreground apps interacting with users~\cite{xiang2018appclassifier}. 
Therefore, traffic traces, i.e., sequences of consecutive frames, are picked out if the current packet rate is larger than a threshold $\gamma$. 
In our system, we empirically set $\gamma$ to be three frames per second.

After obtaining raw traffic traces, we proceed to filter out irrelevant 802.11 frames and extract useful metadata information from data frames. 
Typically, the collected wireless traffic contains three types of 802.11 MAC frames, i.e., control frames, management frames, and data frames. 
The former two types are mainly responsible for medium access control, LAN management, device association, and so on.
Thus, they carry little information about users' app usage behaviors. 
Due to this reason, we decode the common fields of captured frames, filter out management and control frames based on frame types, and retain data frames only. 
A data frame consists of three major parts, i.e., header, body, and trailer.
The header part includes metadata fields, such as frame arrival time, frame length, destination address, and source address.
The trailer contains the frame check sequence to validate the integrity of the entire frame.
These two parts are both in plaintext.
On the contrary, the body part encapsulates payloads that carry upper-layer information, which are encrypted for security concerns.
Since we cannot decrypt the body part in the ciphertext, we only extract the plaintext metadata information in the frame header to represent traffic traces. 

Formally, we assume that a total of $T$ traffic traces are collected in the user profiling phase and $I$ MAC addresses are involved. 
Since users are not likely to use their smartphones all the time, multiple traffic traces may belong to the same address. 
Thus, the $j$-th traffic trace of $i$-th MAC address can be represented as
\begin{align}
    F^j_i = \left\{f^{1}, \cdots, f^m, \cdots, f^{M} \right\}.
\end{align}
In the above equation, $f^m=(t^m,s^m,d^m)$ represents the metadata of the $m$-th frame. 
Therein, $t^m\in \mathbb{R}$ and $s^m\in \mathbb{N}$ are, respectively, the arrival time and packet size of the $m$-th frame. 
$d^m \in  \left\lbrace -1,1 \right\rbrace $ represents downlink and uplink packets, which are transmitted from and to the AP, respectively.
$M \in \mathbb{N}$ stands for the total number of Wi-Fi frames in the traffic trace $F^j_i$.
Moreover, let us denote $\Delta$ as the time duration of $F^j_i$. 
Hence, we have $\Delta=t^M-t^1$.
Finally, the collected traffic dataset $\mathcal{D}$ can be obtained as $\mathcal{D} = \left\{ F^j_i \right\}^{j=1:J_i}_{i=1:I}$, where $\sum_{i=1}^{I} J_i = T$.

\begin{table}[htbp]
    \centering
    \caption{Extracted In-App Action Features.}
    \resizebox{\linewidth}{!}{
    \begin{tabular}{cc}
         \hline
         Tags & Features\\
         \hline
         $p_1$& Total number of frames\\
         $p_2$& Average size of all frames\\
         $p_3$& Average time interval\\
         $p_4$& Uplink and downlink ratio\\
         $p_5$& Kurtosis of frame rate\\
         $p_6$& Skewness of frame rate\\
         $p_7$& Average size of uplink traffic\\
         $p_8$& Average size of uplink traffic (low 20\%)\\
         $p_9$& Average size of uplink traffic (mid 60\%)\\
         $p_{10}$& Average size of uplink traffic (high 20\%)\\
         $p_{11}$& Variance of uplink traffic (low 20\%)\\
         $p_{12}$& Variance of uplink traffic (mid 60\%)\\
         $p_{13}$& Variance of uplink traffic (high 20\%)\\
         $p_{14}$& Average time interval for uplink traffic\\
         $p_{15}$& Average size of downlink traffic\\
         $p_{16}$& Average size of downlink traffic (low 20\%)\\
         $p_{17}$& Average size of downlink traffic (mid 60\%)\\
         $p_{18}$& Average size of downlink traffic (high 20\%)\\
         $p_{19}$& Variance of downlink traffic (low 20\%)\\
         $p_{20}$& Variance of downlink traffic (mid 60\%)\\
         $p_{21}$& Variance of downlink traffic (high 20\%)\\
         $p_{22}$& Average time interval for downlink traffic\\
         \hline
    \end{tabular}
    }
    \label{tab:activity_feature}
\end{table}

\textbf{Multi-Level Feature Extraction.} 
Based on the captured traffic traces, we exploit two overlapping sliding windows to extract multi-level features for recognizing apps and actions.

Due to the differences in functionality and communication logic, different apps will present distinct changing patterns of the size, direction, and interval time among adjacent frames.
Since each traffic trace is a frame sequence, a feature extraction scheme using a sliding window is proposed to fully characterize hidden sequential patterns.
The main idea is to divide each trace into equally sized feature samples with a sliding window. 
Specifically, centered at the $m$-th frame in the traffic trace $F^j_i$, the sliding window with a length of $W_s$ is used to generate an app feature sample $u^m$ as 
\begin{align}\label{eq:application_sample}
u^m = \left\{f^{m-(W_s-1)/2},\cdots,f^m,\cdots,f^{m+(W_s-1)/2}\right\}.
\end{align}
In this way, we can yield a feature sample for each frame in $F^j_i$.
From Eq.~\eqref{eq:application_sample}, we can observe that $u^m$ not only retains metadata information of the $m$-th frame but also preserves contextual features of its neighboring frames, which is sufficient for app recognition.

The next step is to extract fine-grained features about in-app actions.
In general, in-app actions, such as text chat and browsing, are highly correlated with user actions, which are continuous in the time domain.
Therefore, we divide the traffic trace $F^j_i$ into traffic bursts, i.e., segments with a time duration of one second each.
In this way, a total of $N$ traffic bursts can be obtained, where $N\times1 \approx \Delta$, i.e., the time duration of $F^j_i$.
Given the $n$-th traffic burst, we extract 22 statistical features as listed in Table~\ref{tab:activity_feature} and obtain a feature vector $p^n=(p_1^n,\cdots,p_{22}^n) \in \mathbb{R}^{22}$.
Then, the sliding window with a size of $W_a$ is leveraged to generate in-app action feature samples.
Specifically, for the $n$-th traffic burst of $F^j_i$, the action feature sample $v^n$ can be written as 
\begin{align}
v^n = \left\{p^{n-(W_a - 1)/2},\cdots,p^n,\cdots,p^{n+(W_a - 1)/2}\right\}.
\end{align}
Through the above process, fine-grained statistical and contextual information can be extracted for action classification.

\begin{figure}
    \centering
    \includegraphics[width=1\linewidth]{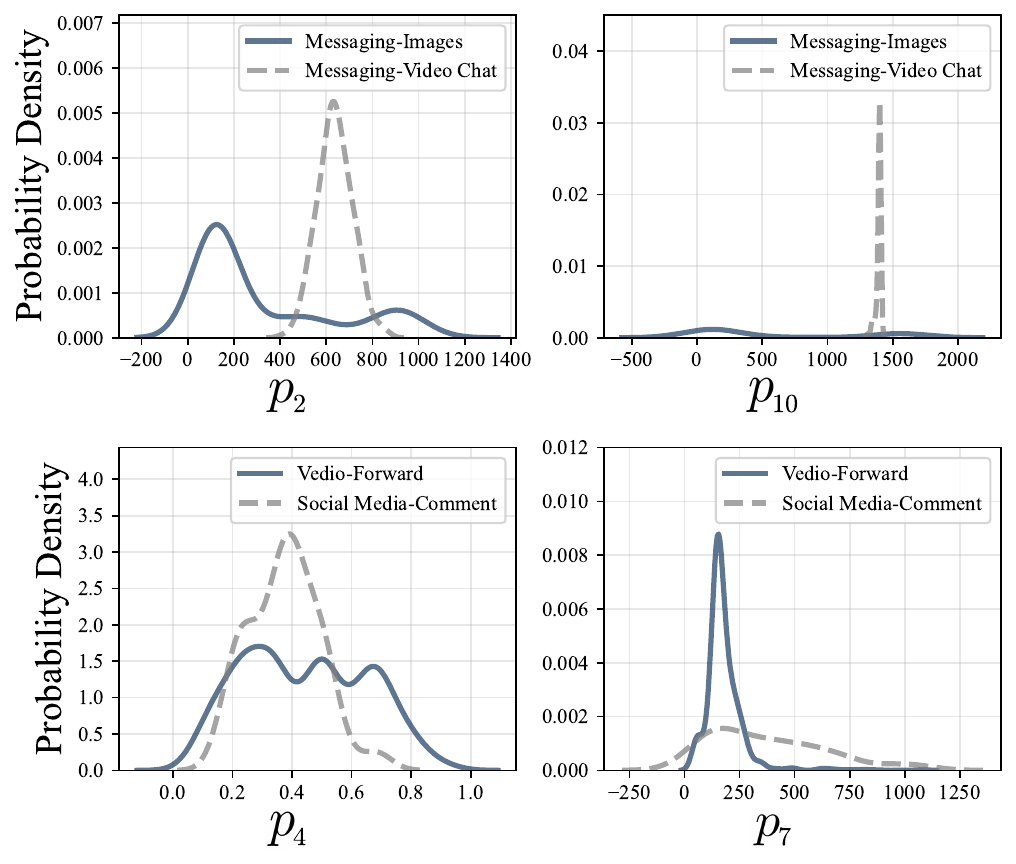}
    \caption{Density distributions of statistical features across different actions.}
    \label{fig:density_distribution}
\end{figure}

To validate the effectiveness of the selected features, we plot probability density distributions of four statistical features, i.e., $p_2$, $p_4$, $p_7$, and $p_{10}$ in Table~\ref{tab:activity_feature}, across different in-app actions. 
The probability density is estimated using kernel density estimation with a Gaussian kernel based on the datasets collected in our experiments.
As Fig.~\ref{fig:density_distribution} shows, each action exhibits a unique feature distribution from the others, even for two actions in the same-category apps, e.g., sending images and video chatting in messaging apps.

\subsection{App and Action Recognition}
Based on the multi-level features, U-Print recognizes which apps of interest are used and which in-app actions are performed by mobile users.
To achieve this goal, we devise an app classifier and a set of action classifiers.
Moreover, an Android app, namely the label recorder, is developed to annotate traffic traces for effective classifier training. 

\begin{figure}
    \centering
    \includegraphics[width=1\linewidth]{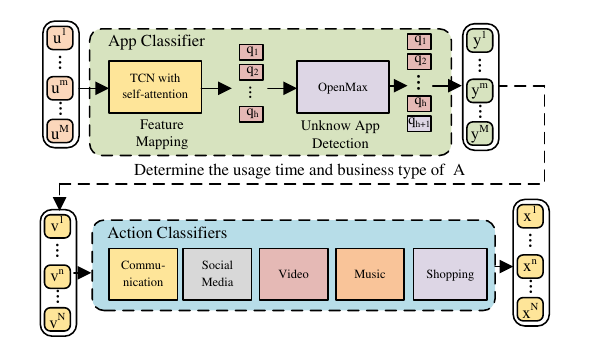}
    \caption{App classifier and action classifiers.}
    \label{fig:Recognitionprocess}
\end{figure}
 
\textbf{App Classifier.} Taking an app feature sample $u^m$ as input, our app classifier $\mathcal{F}_{app}(\cdot)$ outputs a one-hot vector $y^m \in \left\{0,1\right\}^{H+1}$, where 1 indicates the corresponding app and $H$ is the number of apps of interest.  
In addition, the last element in $y^m$ stands for apps that are unseen in model training.
For effective recognition, the app classifier consists of feature mapping and unknown app detection as shown in Fig.~\ref{fig:Recognitionprocess}.

In feature mapping, we develop a temporal convolutional network with a self-attention mechanism to convert an app feature sample $u^m$ into a probability vector $ Q^m = [q^1,\cdots, q^{H}] \in [0,1]^{H} $.
The reasons for adopting TCNs are twofold.
First, TCNs~\cite{bai2018empirical} are composed of a set of dilated causal convolutional layers and can abstract underlying temporal and contextual information from time series data, making them suitable for processing sequences of wireless frames.
Second, compared with traditional networks like recurrent neural networks, TCNs generally have better performance in training speed and scalability.
Moreover, we enhance our TCN model with a self-attention mechanism, which helps the model extract long-term dependencies in the input traffic trace.

In unknown app detection, we leverage an OpenMax~\cite{bendale2016towards} function to extend the $H$-dimensional probability vector $ Q^m $ into a calibrated one $ \hat{Q}^m $ with $H+1$ dimensions to facilitate open-set classification.
Typically, most existing traffic fingerprinting techniques are based on closed-set classifiers~\cite{wang2015know,taylor2016appscanner,atkinson2018your}, which can only handle known categories presented in model training but cannot correctly recognize unseen ones during the testing phase.
In practice, there are millions of apps in various mobile app stores~\cite{appnum}, making it a non-trivial task to collect wireless traffic from all of them.
To deal with this issue, we employ an OpenMax function.
Mathematically, given the app feature sample $u^m$, this function converts the corresponding TCN output $ Q^m $ into $ \hat{Q}^m \in [0,1]^{H+1}$ as
\begin{align}\label{eq:calibrated_vector}
    \hat{Q}^m  = [q^1c^1,\cdots, q^hc^h,\cdots, q^{H} c^{H}, q^{H+1}],
\end{align}
where $q^{H+1}$ is the probability for unknown apps and can be calculated by
\begin{align}\label{eq:probability_unknown}
   q^{H+1} = \sum_{h=1}^{H} q^n (1-c^h).
\end{align}
In the above two equations, $c^h \in [0,1]$ is a confidence weight, indicating the probability that $u^m$ is the $h$-th category. 
The confidence $c^h$ is estimated by measuring the distance between $u^m$'s activation vector, i.e., the feature map of the last fully connected layer in the app classifier, and that of correctly classified training samples of the $h$-th category using the Weibull distribution.
In general, the closer they are, the bigger the $c^h$ is.
By using the OpenMax function, our app classifier can learn the feature space of known apps and detect unknown ones, thus improving its performance in open-world classification.
Using the calibrated probability vector $ \hat{Q}^m $, the app classifier predicts the most likely app $h^m$ as
\begin{align}
    h^m = \left\lbrace 
    \begin{array}{lcl}
    H+1 \: , & \text{if} \: \: q^{H+1} > \delta ; \\
    \arg \underset{{h=1:H}}{\max} \: \hat{Q}^m, \:  & \text{else} .
\end{array}
\right.
\end{align}
Therein, $\delta$ is a threshold. 
When $q^{H+1}>\delta$, the classifier predicts the sample $u^m$ as an unknown class. Otherwise, it can be considered from the app with the highest probability. 
Finally, we can utilize one-hot encoding to transform $h^m$ into the one-hot vector $y^m$.

Toward this end, we can feed all app feature samples of the traffic trace $F^j_i$ into our app classifier $\mathcal{F}_{app}(\cdot)$ and obtain a sequence of app labels as
\begin{align}
    Y^{j}_{i}  = \mathcal{F}_{app}(F^j_i) = \left\{y^1, \cdots, y^m, \cdots, y^M \right\}.
\end{align}
Then, to assign an app prediction to each traffic burst, we divide $Y^{j}_{i}$ into one-second segments based on burst timestamps.
For each segment, the most frequent label is selected as the app label of the corresponding burst.
In this way, we obtain another app label sequence $\hat{Y}^{j}_{i} $ as
\begin{align}
    \hat{Y}^{j}_{i}  = \left\{\hat{y}^1, \cdots, \hat{y}^n, \cdots, \hat{y}^N \right\}.
\end{align}
In addition, to effectively train our app classifier $\mathcal{F}_{app}(\cdot)$, we use the cross-entropy loss to measure the differences between its predictions and the ground-truth labels in the training phase.  

\begin{table}
    \centering
    \caption{40 Selected Apps from the China Android Market and Google Play Store.}
    \setlength{\tabcolsep}{0.3mm}{
    \begin{tabular}{c|c|c}
         \hline
         \textbf{App Category}& \textbf{Apps of Interest} & \textbf{In-App Actions}\\
         \hline
         \textbf{Messaging}&  
         \begin{tabular}{cccc}
              WeChat&QQ&WhatsApp&Telegram\\
              Messenger&Snapchat&Hangouts&Discord\\
         \end{tabular}
         &
         \begin{tabular}{cc}
              Text&Voice Chat\\
              Images&Video Chat\\
         \end{tabular}\\
         \hline
         \textbf{Social Media}& 
         \begin{tabular}{cccc}
              Weibo&Baidu Tieba&Quora&Facebook\\
              Twitter&Red Booklet&Reddit&Instagram\\
         \end{tabular}
         & 
         \begin{tabular}{cc}
              Browsing&Comment\\
              Thumb-up&Share\\
         \end{tabular}\\
         \hline
         \textbf{Video}&
         \begin{tabular}{cccc}
              YouTube&Tiktok&Netflix&Vimeo\\
              Tencent Video&Bilibili&Twitch&iQIYI\\
         \end{tabular}
         & 
         \begin{tabular}{cc}
              Forward&Play\\
              Backward&Next\\
         \end{tabular}\\
         \hline
         \textbf{Music}& 
         \begin{tabular}{c}
              \begin{tabular}{ccc}
              NetEase Cloud&QQ Music&Spotify\\
              \end{tabular}\\
              \begin{tabular}{ccc}
              SoundCloud&Apple Music&Shazam\\
              \end{tabular}\\
              \begin{tabular}{cc}
                  Kugou Music& YouTube Music\\
              \end{tabular}              
         \end{tabular}
         & 
         \begin{tabular}{cc}
              Forward&Play\\
              Backward&Next\\
         \end{tabular}\\
         \hline
         \textbf{Shopping}& 
         \begin{tabular}{cccc}
              Taobao&JD&PDD&Amazon\\
              eBay&Walmart&Rakuten&Suning\\
         \end{tabular}
         & 
         \begin{tabular}{cc}
              Search&Browsing\\
              Cart&Checkout
        \end{tabular}\\
        \hline
    \end{tabular}
    \label{tab:applist}
    }
\end{table}

\textbf{Action Classifiers.} 
As aforementioned, millions of mobile apps could run on smartphones, making it an impossible task to build a dedicated action classifier for each app.
For this reason, we select 40 high-ranking mobile apps from both the China Android Market and Google Play Store as the apps of interest, which are presented in Table~\ref{tab:applist}.
The chosen apps span multiple categories, including messaging, social media, video, music, and shopping, providing a comprehensive representation of daily smartphone usage. 
In each category, the common actions are summarized in Table~\ref{tab:applist}.

In this way, a total of five action classifiers are built. 
Without loss of generality, let us denote an action classifier as $\mathcal{F}_{act}(\cdot)$.
Given the action feature sample $v^n$, it outputs a one-hot vector $x^n \in \left\{0,1\right\}^{G+1}$, where $G$ indicates the total number of common actions.
For simplicity, $\mathcal{F}_{act}(\cdot)$ has the same architecture with $\mathcal{F}_{app}(\cdot)$.
In the inference stage, U-Print decides to use which action classifier based on predicted app labels.
Finally, given the traffic trace $F^j_i$, U-Print outputs a sequence of action labels as
\begin{align}
    X^{j}_{i}  = \mathcal{F}_{act}(F^j_i) = \left\{x^1, \cdots, x^n, \cdots, x^N \right\}.
\end{align}

To this end, according to $\hat{Y}^{j}_{i}$ and $X^{j}_{i}$, we can obtain a user operation sequence $B_i^j$ from the traffic trace $F^j_i$ as
\begin{align}
    B_i^j = \left\{b^1,\cdots,b^n,\cdots,b^N\right\},
\end{align}
where $b^n=\left\{y^n, x^n \right\}\in \left\{0,1\right\}^{H+G+2}$ and is the concatenation of $y^n$ and $x^n$.
It represents the user operation of the $n$-th traffic burst, which uniquely indicates which app is used and which in-app action is performed.

\textbf{Label Recorder.} Considering that our app classifier outputs a prediction for each MAC frame, a dataset with frame-level annotation is necessary for model training.
However, existing traffic fingerprinting approaches mainly rely on manual annotation for each traffic trace, which cannot accurately label MAC frames due to the quick and frequent switching among different apps and actions by users in reality.
To solve the labeling problem, we design a label recorder, an Android app that runs in the background on users' smartphones and collects log information relative to user interaction events.
When running on the smartphone, our label recorder captures information about the foreground app as well as the time and location of screen taps and saves it as an interaction log in a txt format on the smartphone.
The label recorder works based on the TouchEventAccessibilityService class, where the currentTimeMillis, getPackageName, and getSource methods are leveraged to acquire the screen tapping time, the name of the foreground app, and the screen tapping location.
Thus, the label recorder allows us to obtain the interaction log as
\begin{align}
    L = \left\{ \cdots,l^{r-1},l^r,l^{r+1},\cdots \right\},
\end{align}
where $l^r=(t^r,b^r,d^r)$ represents the $r$-th record. 
Here, $t^r$, $b^r$, and $d^r$ denote the tapping time, the name of the foreground app, and the tapping location, respectively. 

\begin{figure}[t]
\centering
\subfloat[]{\includegraphics[width=0.48\linewidth]{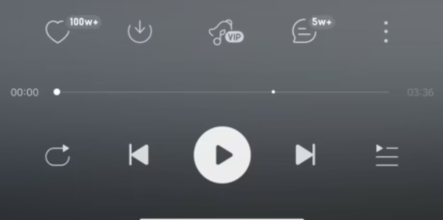}
\label{Location}}
\hfil
\subfloat[]{\includegraphics[width=0.48\linewidth]{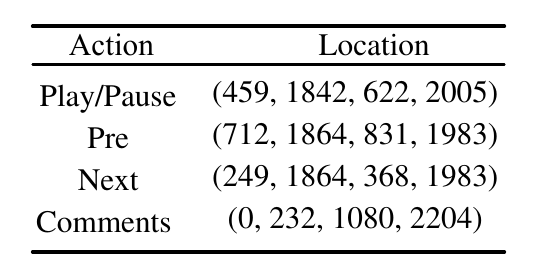}
\label{Mapping table}}
\caption{The relationship between button locations in the app UI and in-app actions. (a) App UI. (b) Mapping table.}
\label{fig:UIinterface}
\end{figure}

During data collection, we run the label recorder on the user's smartphone to generate the interaction log and exploit Wi-Fi sniffing tools to capture corresponding traffic traces.
To annotate traffic frames with specific apps, we exploit the temporal correlation between user interaction and the interaction log.
To do this, we rely on the time and app name records to segment the period of the log into many intervals, during which the same app runs in the foreground. 
Subsequently, the traffic frame within the same interval is labeled with the corresponding app name.
To annotate traffic traces with action labels, we leverage the spatial correlation between in-app actions and the app user interface (UI).
Typically, each app has a fixed UI, and in-app actions are activated by the click of some predefined locations on the screen.
This implies that there is a correlation between the tapping location and in-app actions, allowing us to infer performed actions from tapping location records.
To demonstrate that, a music player's UI is presented in Fig.~\ref{fig:UIinterface}~(a), and the buttons for in-app actions, such as play/pause, and previous/next song, are located at fixed UI locations. 
Based on this observation, we first establish a mapping table about the correlation between the click location and in-app actions for each app of interest.
Fig.~\ref{fig:UIinterface}~(b) illustrates the correlation table for the above music player.
Subsequently, we label one traffic burst with a specific action if its correlated location is pressed within the burst time.
Moreover, if no targeted action is detected, the burst is labeled with an unknown class.
In this way, we can obtain labeled traffic traces to train our app and action classifiers.
Note that the label recorder is installed only on the controlled users' smartphones for training the app and action classifiers.
Once trained, they are deployed to infer apps and actions on the victim users' devices.

\subsection{User Profiling and Identification} 
Due to the wide adoption of the MAC randomization strategy, smartphones periodically replace their original MAC addresses with pseudonyms, i.e., unlinkable names, when communicating with the AP. 
For instance, most Android smartphones change their MAC addresses once a day~\cite{macrandom}.
Owing to this reason, one user's device would have multiple randomized MAC addresses throughout the profiling process, rendering it infeasible for U-Print to identify smartphone users based on their temporary MAC addresses.
To tackle this challenge, we observe that mobile users are tempted to install different apps based on their interests and even exhibit unique usage patterns in the same app.
Moreover, such usage patterns are likely to be similar in consecutive days~\cite{li2020extent}.
In this way, the operation sequences from the same user would be very similar.
Based on this observation, we use the simple and scalable k-means clustering algorithm on all app and action predictions to build user profiles for victim users and further recognize them. 
Specifically, the K-Means algorithm is suitable for our task, where different users show unique usage patterns, and the number can be reasonably estimated.
Moreover, its clustering results are interpretable and can be considered as user profiles.

\begin{figure}
    \centering
    \includegraphics[width=1\linewidth]{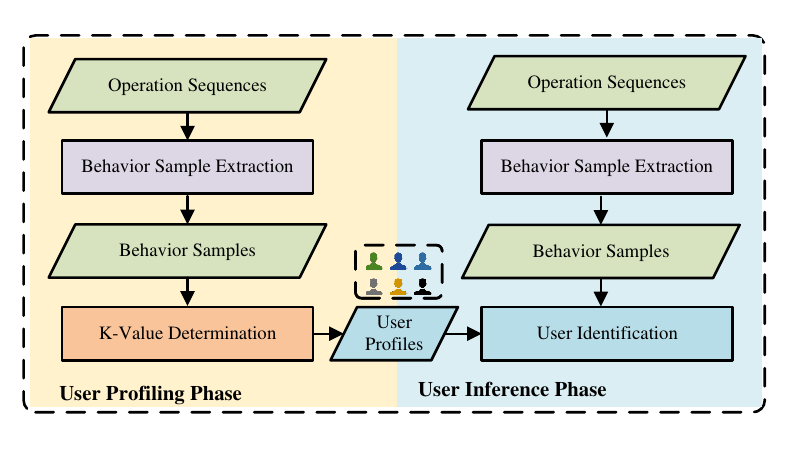}
    \caption{The workflow of user profiling and identification.}
    \label{fig:useridentification}
\end{figure}

To achieve this goal, we need to obtain behavior samples representing users' usage habits and determine the number of users for building user profiles.
After that, the user profiles can be used for user identification in the inference phase.
The workflow of user profiling and identification is shown in Fig.~\ref{fig:useridentification}.

\textbf{Behavior Sample Extraction.} 
After obtaining operation sequences, we proceed to extract behavior samples, i.e., the basic units for user profiling and identification. 
To do this, we simplify an operation sequence $B_i^j$ via a linear scanning method, which converts its consecutive identical labels into one.
In this way, we can obtain a behavior sequence $\hat{B}_i^j$ as
\begin{align}
    \hat{B}_i^j = \left\{\hat{b}^1,\cdots,\hat{b}^e,\cdots,\hat{b}^E\right\},
\end{align}
where $E \leq N$ and $\hat{b}^e$ is taken from $B_i^j$.
The reason behind this operation is that the lasting time of some actions, such as browsing and texting, is different each time, making raw app and action predictions less effective in characterizing users' usage patterns. 
By simplifying labels that repeat in a row, U-Print can remove redundant information in operation sequences and retain the order information of different operations.

The next step is to extract fixed-length behavior samples from the behavior sequence $\hat{B}_i^j$ to reflect the unique usage patterns of different users.
In the user profiling phase, user profiles should contain behavior samples as much as possible.
For this purpose, we employ a sliding window technique.
Specifically, we use a sliding window with a length of $W_b$ and a stepping size of 1 to extract behavior samples from the $\hat{B}_i^j$ and obtain a set of behavior samples as $\mathcal{Z}_i^j = \left\{z^1, \cdots, z^e, \cdots, z^{E-W_b+1}  \right\}$.
In $\mathcal{Z}_i^j$, each element $z^e$ can be represented as
\begin{align}
    z^e = \left\{\hat{b}^e,\cdots,\hat{b}^{e+W_b-1}\right\}.
\end{align}
The above sliding window technique can generate multiple overlapping subsequences of one behavior sequence, thus increasing the diversity of training data in the user profiling phase.

\begin{figure}
    \centering
    \includegraphics[width=0.9\linewidth]{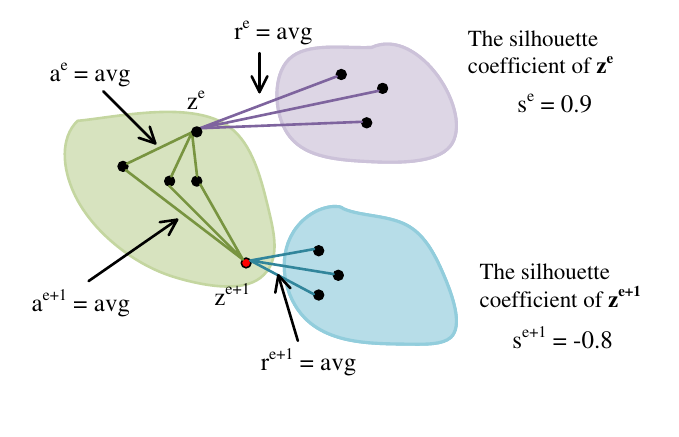}
    \caption{Illustration of silhouette coefficients. $z^e$ is the correctly clustered sample with a score close to 1, and $z^{e+1}$ is the incorrectly clustered sample with a score close to -1.}
    \label{fig:sc}
\end{figure}

\textbf{K Value Determination.} 
In the k-means clustering, the $k$ value, i.e., the number of clusters, needs to be specified at the beginning.
However, during the profiling phase, the number of smartphone users in the targeted area is unknown to the attacker.  
To deal with this issue, we use the silhouette coefficient method to determine the $k$ value. 

First, the SCM specifies the value of $k$ in a range of $\left\{1,\cdots, K \right\}$ and leverages the k-means clustering to divide all samples into $k$ clusters.
In our system, we set $K$ to be the number of all sniffed MAC addresses.
Then, the average silhouette coefficient of all samples $S_k$ can be calculated.
Specifically, the silhouette coefficient of one behavior sample $z^e$ is represented as 
\begin{align}
    s^e = (r^e-a^e)/\max\left\{r^e,a^e\right\},
\end{align}
where $a^e$ represents $z^e$'s average distance to all other samples in the same cluster, i.e., suggesting the similarity of one sample to its cluster, and $r^e$ represents $z^e$'s distance to all samples in the nearest cluster, i.e., indicating the similarity of one sample to other clusters.
To measure distances, we use the Hamming distance to calculate the difference between two samples because our behavior samples are binary vectors. 
Concretely, given two behavior samples, we first compare their MAC addresses.
If the addresses are identical, we set their distance to zero.
If not, we employ the Hamming distance to measure their difference. 
In this way, the average silhouette coefficient of $ \left\{ \mathcal{Z}_i^j \right\}^{j=1:J_i}_{i=1:I}$ is obtain by 
\begin{align}
    S_k=\frac{1}{Z}\sum_{i,j} \sum_{e}s^e ,
\end{align}
where $Z$ is the total number of behavior samples and $S_k \in (-1, 1)$. 
As illustrated in Fig.~\ref{fig:sc}, the sample $z^e$, which is correctly clustered, has a silhouette coefficient $s^e = 0.9$, while sample $z^{e+1}$, which is misclassified, has a silhouette coefficient $s^{e+1} = -0.8$.
This demonstrates that the closer the silhouette coefficient is to 1, the smaller the intra-cluster difference and the larger the inter-cluster difference, indicating a better clustering effect.

Next, the SCM calculates the silhouette coefficients under different values of $k$.
We select the optimal $k$ value that has the maximum silhouette coefficient as 
\begin{align}
    k_{op} = \arg \underset{k}{\max}\left\{S_1,S_2,\cdots,S_{K}\right\}.
\end{align}
In this way, the user number in the targeted area can be determined.
 
\textbf{User Identification.} In the user profiling phase, we apply the k-means clustering to group all behavior samples based on the user number $k_{op}$ and obtain their profiles as
\begin{align}
    \mathcal{P} = \left\{\mathcal{\hat{Z}}_1, \cdots, \mathcal{\hat{Z}}_{k_{op}}\right\}.
\end{align}
Therein, $\mathcal{\hat{Z}}_k$ consists of all behavior samples of $k$-th user.
The centering samples of all clusters can be obtained as $\mathcal{C} = \left\{\mu_1, \cdots, \mu_{k_{op}}\right\}$.
Thus, a user profile contains the history of user-smartphone interactions, which can be further exploited to speculate on the user's age, gender, health status, sexual orientation, and other private information~\cite{wang2015know}.

In the inference phase, given a new behavior sample $z_t$, U-Print calculates its distances to all centering samples in $\mathcal{C}$ using the Hamming distance.
Then, it selects the nearest cluster as
\begin{align}
    k_t = \arg \underset{k}{\min} \; \text{Hamming}(z_t,\mathcal{C}),
\end{align}
which means the behavior sample $z_t$ is incurred by $k_t$-th user.
In the long run, our system can exploit samples collected in recent days to update user profiles to deal with behavioral pattern shifts.
In this way, U-Print exploits behavioral continuity in user-smartphone interactions to associate wireless traces without static hardware identifiers or special Wi-Fi frames, and it can recognize who is using which mobile app with which in-app action on her/his smartphone.

\section{Evaluation}

\subsection{Evaluation Methodology}

\begin{figure}
    \centering
    \includegraphics[width=1\linewidth]{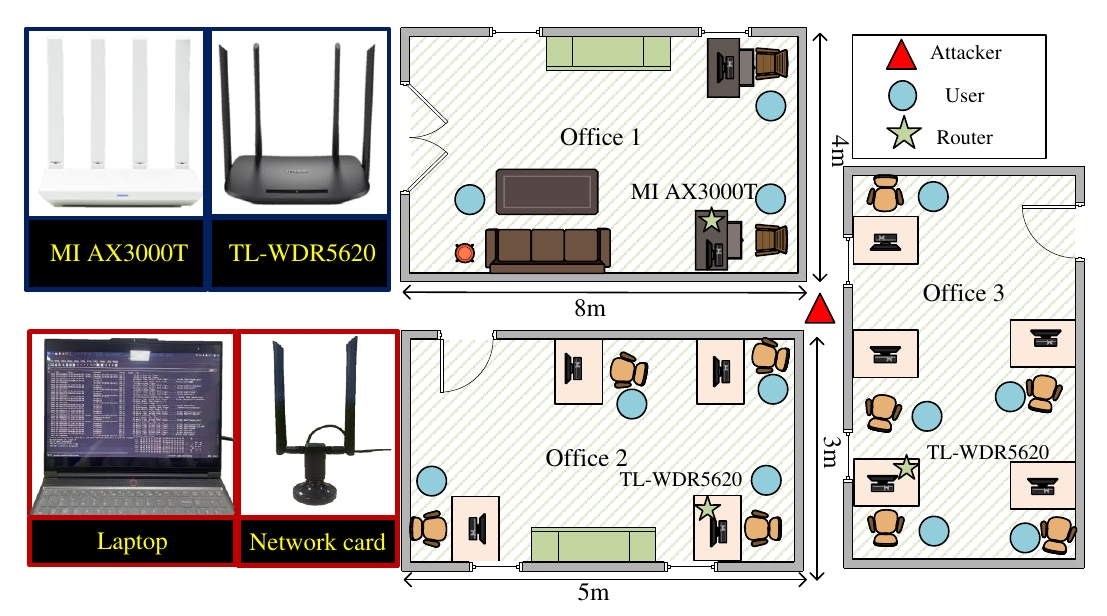}
    \caption{Experimental setup.}
    \label{fig:setting}
\end{figure}

\textbf{Implementation.} 
We implemented U-Print using a Lenovo laptop connected to a Kali dual-band network card, as shown in Fig.~\ref{fig:setting}.
The laptop runs Kali Linux and is powered by an Intel Core i7-12700H.
It is equipped with an NVIDIA GeForce RTX 3090 GPU with 24 GB of VRAM and 16 GB of system memory.
The working frequency range of the network card is 2.4 GHz–2.4835 GHz and 5.125 GHz–5.825 GHz.
We set the network card to the monitoring mode using Aircrack-ng to search all active Wi-Fi APs.
Then, we used Wireshark to collect MAC layer traffic data from the targeted AP and filtered out management and control frames.

\textbf{Data Collection and Datasets.} 
To the best of our knowledge, no public datasets can be found in the relevant work. 
Thus, we built three custom datasets for evaluation.
To generate labeled samples in the three datasets, we collected raw wireless traffic and simultaneously recorded actual user activities and user identities.
Then, each traffic segment was denoted with actual labels based on synchronized timestamps.

First, we constructed Dataset A as a closed-world dataset to evaluate U-Print's performance on app and in-app action classification, where all testing samples came from apps and actions known during training.
Before traffic collection, we installed 40 apps listed in Table~\ref{tab:applist} on an MI 8 UD smartphone that was connected to a TL-WDR5620 Wi-Fi router for Internet access.
During traffic collection, we first ran our designed label recorder, an Android app, on the MI smartphone to record the user-device interaction log for accurate data labeling.
Then, one user randomly performed defined and undefined actions on apps of interest to generate Wi-Fi traffic.
Next, we sniffed the corresponding wireless traffic data from the MI smartphone using the Kali network card.
We collected 15 traffic instances for each app, with each instance lasting 10 minutes, and obtained MAC layer traffic with a total of 100 hours.
In this way, about 360K traffic samples are included in Dataset A.

Second, we constructed Dataset B as an open-world dataset to evaluate U-Print's performance on app and in-app action classification, where the testing samples include apps and actions that were unseen during training.
In addition to the apps included in Table~\ref{tab:applist}, we collected traffic instances of 10 apps, such as Youku and Zhihu, in the same way. 
Therefore, a total of 450K traffic samples are included in Dataset B.

Third, we constructed Dataset C as a user profiling and identification dataset to evaluate U-Print's performance on user identification. 
This dataset contains traffic samples from multiple users performing various app activities over time.
For this purpose, we recruited a total of 12 volunteers and collected wireless traffic from three office rooms on our campus, as shown in Fig.~\ref{fig:setting}.
Specifically, there were three users in the first office, and all of them connected their smartphones to an MI AX3000T wireless router. 
Moreover, there were four and five users, respectively, in the second and third offices, where a TL-WDR5620 Wi-Fi router was mounted to provide wireless connectivity in each environment.
The ages, genders, and smartphones of 12 volunteers are presented in Table~\ref{tab:userinfo}.
When entering each office, we required each volunteer to connect her/his smartphone to the corresponding router at the beginning and then report the smartphone's MAC address.
Then, the sniffer was tuned to the channel used by this MAC address and received its wireless traffic.
During traffic collection, each volunteer ran various apps and performed in-app actions freely.
We conducted traffic collection from three offices for one week.
Based on the raw traffic data, we segmented out effective traffic data that lasts 25 hours in total, yielding about 8K traffic traces in Dataset C.
On average, each volunteer contributed traffic traces of 5 hours in one day.
Note that our label recorder was only used when constructing the former two datasets for evaluating app and action classification.
The samples in Dataset C was annotated with user IDs with manual labeling, thus both Android and iOS devices were involved in the data collection.

\begin{table}
    \centering
    \caption{Details of 12 Volunteers in Three Environments.}
    \begin{tabular}{ccccc}
    \hline
        UID & Scenario & Age & Gender & Phone \\
        \hline
         User 1& Office 1 & 23 & Male & iPhone 11 pro \\
         User 2& Office 1 & 52 & Male & OPPO Reno11 \\
         User 3& Office 1& 51 & Female & OPPO K11 \\
         User 4& Office 2 & 25 & Male & iPhone 12\\
         User 5& Office 2 & 24 & Female & iPhone 11\\
         User 6& Office 2 & 23 & Male & iPhone 14\\
         User 7& Office 2 & 24 & Male & Huawei nova 12 \\
         User 8& Office 3 & 24 & Male & iPhone 15 pro \\
         User 9& Office 3 & 23 & Male & Oneplus ace3\\
         User 10& Office 3 & 22 & Female & Huawei mate40\\
         User 11& Office 3 & 23 & Male & iPhone 13\\
         User 12& Office 3  & 24 & Female & iPhone 12\\
         \hline
    \end{tabular}
    \label{tab:userinfo}
\end{table}

\textbf{Training and Testing.} We implemented user profiling and identification on Python 3.8 and PyTorch 2.1.0.
In the training phase, we trained a total of six TCN models, including one app classifier and five action classifiers.
Specifically, we divided Dataset A into a training set and a testing set with a ratio of 6:4 and then used the training set to train six classifiers. 
For each classifier, we set its convolutional channels to 64, the kernel size to 3, the dropout rate to 0.2, and the learning rate to 0.001. 
Moreover, we used the cross-entropy loss and adopted the Adam optimizer for model training.
In this condition, all classifiers were trained with 20 epochs.
Additionally, we divided Dataset C into a training set and a testing set with a ratio of 6:4. 
Then, we used the training set of Dataset C to determine the number of smartphone users in each environment.

In the testing phase, we evaluated all classifiers in closed-world (CW) and open-world (OW) settings. 
In the closed-world setting, the mobile apps used in the testing phase were the same as those in the training phase.
However, in the open-world scenario, the apps involved in the testing phase included unseen ones in the training phase.

\textbf{Evaluation Metrics.} We used the following metrics to evaluate the performance of our system. 
\begin{itemize}
    \item \textbf{Accuracy.} It is defined as the ratio of the number of samples that are correctly classified to the total number of samples.
    \item \textbf{F1 score.} The F1 score is the harmonic mean of precision and recall, which comprehensively reflects model classification performance.
\end{itemize}

\subsection{Experimental Results}

\textbf{Overall Performance.} 
First, we presented the overall performance of U-Print.
For this purpose, we used Dataset A and Dataset B to test the performance of U-Print on app and action classification in closed-world and open-world settings, respectively, and used Dataset C to evaluate its user identification performance.
As shown in Table~\ref{tab:overall performance}, U-Print achieved an accuracy of over 96\% and an F1 score of over 0.96 for recognizing mobile apps and in-app actions in the closed-world setting.
When it comes to the open-world app and action classification, our system obtained an accuracy of over 86\% and an F1 score of over 0.85 in the open-world setting.
The performance decrease is due to the introduction of unseen apps in Dataset B. 
Nevertheless, U-Print achieved a user identification accuracy of 98.4\% and an F1 score of 0.983 in the open-world setting, suggesting the high effectiveness of our system in smartphone user fingerprinting.
\begin{table}
    \centering
    \caption{Overall Performance of U-Print.}
    \begin{tabular}{ccc}
    \hline
         & Accuracy & F1 Score\\
         \hline
         App Classification (CW)& 98.7\% & 0.974\\
         Action Classification (CW)& 96.8\% & 0.968\\
         App Classification (OW)& 87.6\% & 0.923\\
         Action Classification (OW)& 86.1\% & 0.857\\
         User Identification (OW)& 98.4\% & 0.983\\
         \hline
    \end{tabular}
    \label{tab:overall performance}
\end{table}

\textbf{Impact of Sliding Window Size.} 
Next, we showed the impact of sliding window size on multi-level feature extraction.
As mentioned in Section III, we adopt a sliding window approach to extract app-level and action-level features. 
The choice of window sizes may impact the performance of U-Print on app and action classification.
Theoretically, the larger the window size, the better the performance is achieved. 
In our experiment, we varied the window size from 10 to 50 in the task of app classification.
In each setting, app-level features were extracted from this window and fed into our app classifier.
Similarly, we varied the window size from 2 to 10 seconds when performing action classification.

\begin{figure}
    \centering
    \includegraphics[width=1\linewidth]{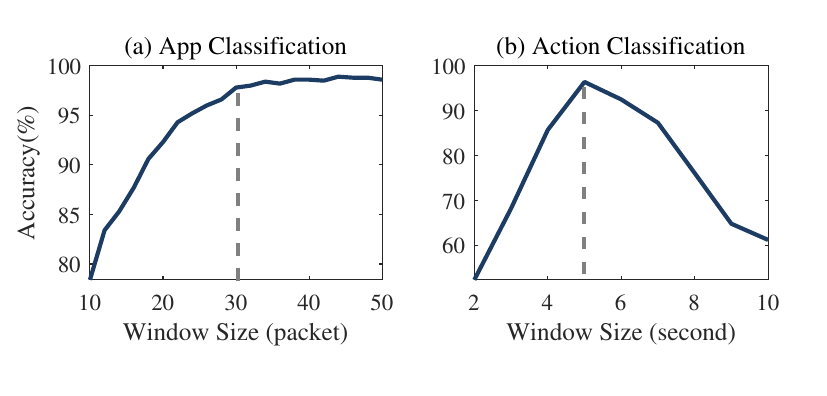}
    \caption{App and action classification based on different sliding windows.}
    \label{fig:differentwindow}
\end{figure}

As depicted in Fig.~\ref{fig:differentwindow}~(a), when the window size increased from 10 to 30, a rapid performance improvement was observed.
However, when it exceeded beyond 30, the accuracy remained around 98\%. 
The above observation indicates that smaller windows contain fewer contextual features, making them more susceptible to noise and leading to lower classification accuracy. 
Conversely, larger windows encompass more Wi-Fi frames, providing richer app-level features and resulting in higher classification accuracy.
Based on the above result, we set the window size to 30 frames in app-level feature extraction.
As shown in Fig.~\ref{fig:differentwindow}~(b), an accuracy increase was observed when the window size increased from 2 to 5. 
But when it rose from 5 to 10, the classification accuracy decreased significantly. 
This may be due to that the majority of in-app actions can be completed within 5 seconds. 
When the window size exceeded 5 seconds, traffic patterns from other actions was introduced, leading to an accuracy decrease.
From the above observations, we set the window size to 5 seconds in action-level feature extraction.

\begin{figure}
    \centering
    \includegraphics[width=1\linewidth]{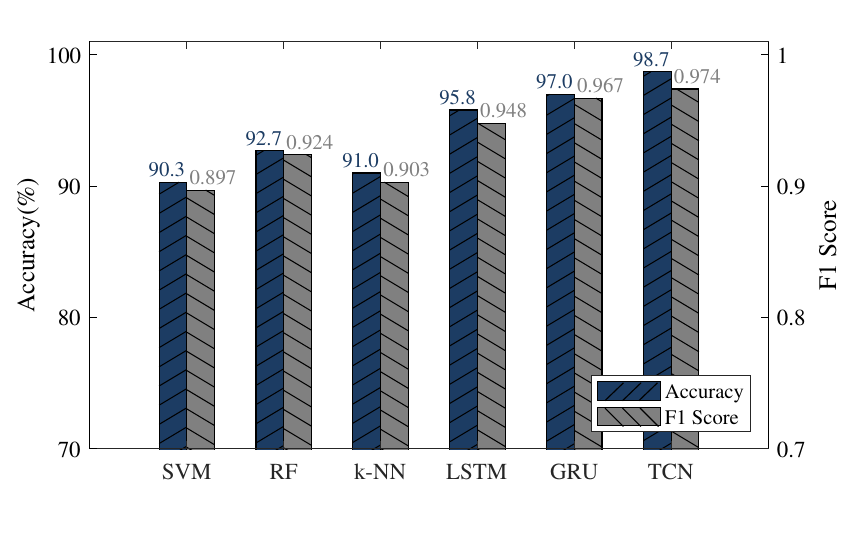}
    \caption{Performance of different classifiers.}
    \label{fig:differentclassfier}
\end{figure}

\textbf{Impact of TCN.} 
Then, we compared the adopted TCN with other classification models.
To achieve this goal, we built five classification models and evaluated them on Dataset A.
Specifically, our baseline classification models included gated recurrent unit (GRU), long short-term memory (LSTM), support vector machines (SVM), random forest (RF), and k-nearest neighbors (k-NN). 
In the GRU and LSTM, we set the number of hidden layers to 2, the layer size to 256, and the learning rate to 0.001. 
Moreover, we chose the radial basis function as the SVM kernel. 
We set the number of estimators to 100 in the RF model and configured the number of nearest neighbors to 10 in the k-NN model.
Fig.~\ref{fig:differentclassfier} presents a performance comparison of different classifiers for app identification.
The prediction results were based on wireless traffic collected from the closed-world dataset.
For each model, we computed the average classification metrics across all 40 apps to ensure a fair and comprehensive evaluation.
We can observe that the accuracy of each model went beyond 90\%. 
Notably, TCN, GRU, and LSTM achieved a classification accuracy of more than 95\% and outperformed the other three models, indicating that deep learning models are better at app classification based on wireless traffic.
Among them, the TCN achieved the highest accuracy rate of 98.7\% and the best F1-score of 0.974.
This is because our TCN consists of a set of dilated causal convolutional layers with a self-attention scheme for effectively abstracting hidden temporal and contextual information from time-series wireless traffic.

\begin{figure}
    \centering
    \includegraphics[width=1\linewidth]{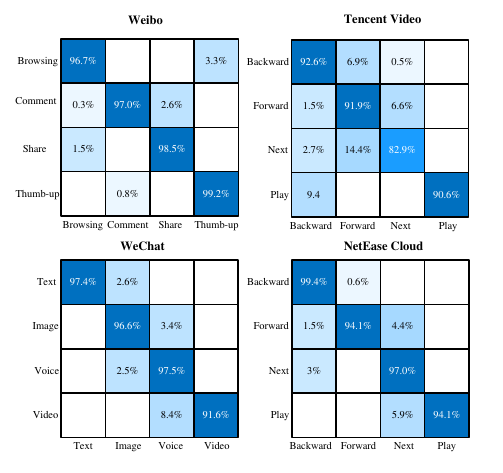}
    \caption{Performance of action classification on WeChat, NetEase CloudMusic, Tencent Video, and Weibo.}
    \label{fig:differentaction}
\end{figure}

Furthermore, we presented the TCN's performance on in-app action classification.
Fig.~\ref{fig:differentaction} showcases its performance on four mobile apps, i.e., Weibo, Tencent Video, WeChat, and NetEase CloudMusic.
The in-app action data were derived from the closed-world dataset.
For each app, we defined and labeled four distinct in-app actions that represent common user behaviors.
Through the confusion matrix of Weibo, we observed that the classification accuracy of the four actions reached more than 96\%.
Similar observations were found in WeChat and NetEase Cloud.
However, when it comes to Tencent Video, the overall accuracy of the four actions was relatively low.
The reason may be that these four common actions all trigger a large amount of downlink traffic, which makes their traffic patterns similar and renders our classifier hard to discriminate.
Despite that, our action classifier still achieved an accuracy of over 90\% for most in-app actions.
The experimental results demonstrate the effectiveness of our TCN in recognizing fine-grained actions.

\begin{figure}
    \centering
    \includegraphics[width=1\linewidth]{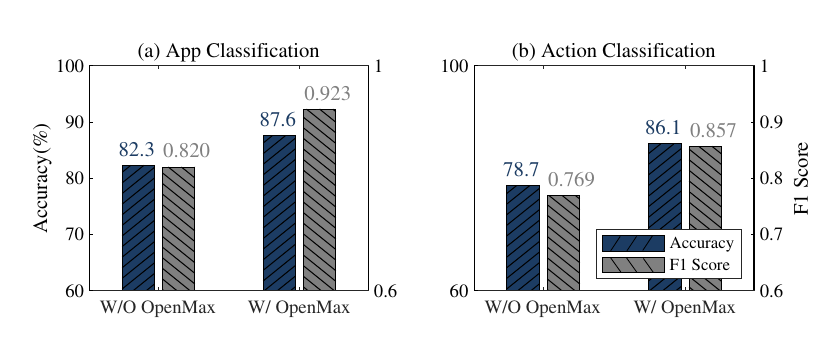}
    \caption{The impact of OpenMax on app and action classification. }
    \label{fig:openmax}
\end{figure}

\textbf{Impact of OpenMax.} 
Furthermore, we demonstrated the effectiveness of the OpenMax function in the open-world app and action classification. 
For this purpose, we built a set of TCN classifiers that are without OpenMax functions and tested them on Dataset B, which contains feature samples from both selected and unseen apps in Table~\ref{tab:applist}.
Fig.~\ref{fig:openmax} depicts the classification performance of U-Print in different settings.
Note that, in open-world tasks, the samples from unseen apps or actions are correctly classified if they are predicted to be unknown.
It can be observed that without the help of OpenMax functions, U-Print achieved a low classification performance, where the app classifier obtained an accuracy of 82.3\% and an F1 score of 0.82, and the action classifiers had an accuracy of 78.7\% and an F1 score of 0.769.
The low classification performance is due to the fact that traditional TCNs cannot deal with unseen samples, which are mistakenly classified into known categories in the testing phase.
However, with the OpenMax function, the accuracy and F1 score of U-Print increased to 87.6\% and 0.923 in app classification, respectively.
The performance boost is attributed to the ability of the OpenMax function to learn the distribution of known samples and discriminate unknown ones.
Similar results can be observed in the task of action classification.
These observations demonstrate that the OpenMax function is effective in recognizing unseen apps and actions during model testing, thus improving the overall performance of U-Print in the open-world environment.

\begin{figure}
    \centering
    \includegraphics[width=1\linewidth]{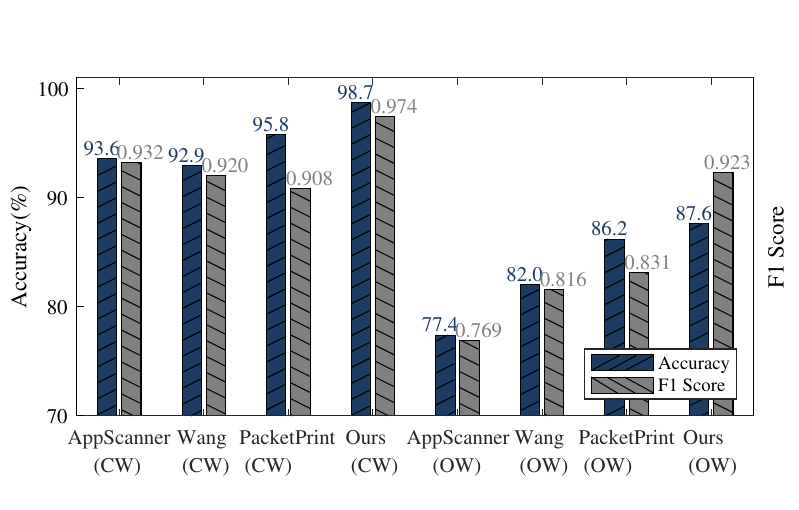}
    \caption{Performance of our system and baselines on app classification.}
    \label{fig:baselines}
\end{figure}

\textbf{Comparison with Baselines.} 
To further illustrate U-Print's advantages in app classification, we built three baseline models, i.e., AppScanner~\cite{taylor2016appscanner}, Wang~\cite{wang2015know}, and PACKETPRINT~\cite{li2022packet}.
Fig.~\ref{fig:baselines} illustrates the app classification performance of our system and three baselines in the closed world and open world, respectively.
To obtain these results, we averaged the classification metrics across all used apps.
It can be clearly observed that our system consistently outperformed three baselines in the two settings.
Especially in the open world, U-Print achieved an F1 score that is 0.15 higher than AppScanner, 0.11 higher than Wang's approach, and 0.09 higher than PACKETPRINT.
This is because U-Print adopts an OpenMax function that can effectively recognize unseen mobile apps in the training phase.
In addition, the performance of our system in the closed world was also higher than that of the three baselines.
This is because U-Print extracts not only statistical features but also fine-grained contextual features, enabling more effective app classification.

\begin{figure}
    \centering
    \includegraphics[width=1\linewidth]{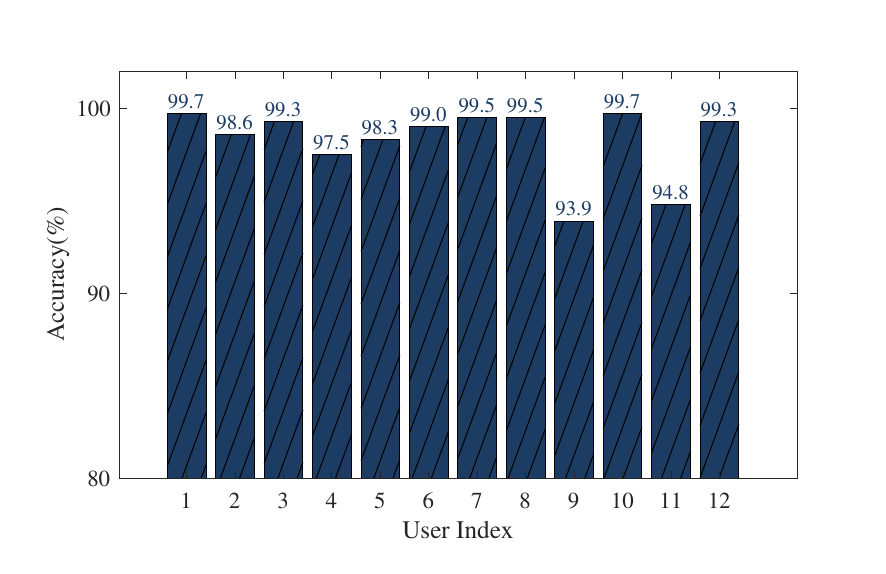}
    \caption{Identification performance per user.}
    \label{fig:allusers}
\end{figure}

\begin{table}[ht]
\centering
\caption{Performance of App Classification.}
\begin{tabular}{l|l c c}
\hline
\textbf{App Category} & \textbf{App} & \textbf{Accuracy}  & \textbf{F1-score} \\
\hline
\multirow{8}{*}{\textbf{Messaging}} & WeChat & 0.977  & 0.973 \\
&QQ & 0.977 & 0.973 \\
&WhatsApp & 0.999 & 0.999 \\
&Telegram  & 0.961  & 0.979 \\
&Messenger & 0.999  & 0.989 \\
&Snapchat  & 0.989  & 0.998 \\
&Hangouts & 0.987  & 0.988 \\
&Discord & 0.987  & 0.989\\
\hline
\multirow{8}{*}{\textbf{Social Media}} & Weibo & 0.963  & 0.979 \\
&Baidu Tieba & 0.951 & 0.967 \\
&Quora & 0.999 & 0.989 \\
&Facebook & 0.989  & 0.969\\
&Twitter & 0.989 & 0.999 \\
&Red Booklet & 0.979  & 0.9998 \\
&Reddit & 0.988  & 0.984 \\
&Instagram & 0.975  & 0.976 \\
\hline
\multirow{8}{*}{\textbf{Video}} & Youtube & 0.962  & 0.969 \\
&Tiktok & 0.989 & 0.939 \\
&Netflix & 0.982 & 0.980 \\
&Vimeo & 0.997 & 0.998 \\
&Tencent Video & 0.960 & 0.963 \\
&Bilibili & 0.982 & 0.980 \\
&Twitch & 0.973 & 0.985 \\
&iQIYI & 0.983 & 0.957\\
\hline
\multirow{8}{*}{\textbf{Music}} &  NetEase Cloud & 0.958 & 0.977 \\
&QQ Music & 0.984 & 0.981 \\
&Spotify & 0.999 & 0.979 \\
&Sound Cloud & 0.999 & 0.999 \\
&Apple Music & 0.985 & 0.987 \\
&Shazam & 0.998 & 0.989 \\
&Kugou Music & 0.987 & 0.988 \\
&YouTube Music & 0.993 & 0.996 \\
\hline
\multirow{8}{*}{\textbf{Shopping}} &Taobao & 0.988 & 0.999 \\
&JD  & 0.988 & 0.984 \\
&PDD & 0.999 & 0.989 \\
&Amazon & 0.989 & 0.989 \\
&eBay & 0.987 & 0.988 \\
&Walmart & 0.977 & 0.983 \\
&Rakuten & 0.983 & 0.981 \\
&Suning & 0.989 & 0.999 \\
\hline
\multirow{1}{*}{\textbf{Average}} & --- & 0.987 & 0.974 \\
\hline
\end{tabular}
\label{tab:closedworld_app}
\end{table}

\textbf{Performance of App Classification.} 
Table~\ref{tab:closedworld_app} summarizes the app classification performance of U-Print on Dataset A, which involves 40 commonly-used mobile apps.
U-Print achieved a high classification performance among all apps, with an average accuracy of 98.7\% and an average F1-score of 97.4\%. 
Especially, it presented near-perfect classification results for many applications, such as WhatsApp, Messenger, Snapchat, and SoundCloud, indicating that their communication patterns are highly distinctive.
The above results demonstrate the high effectiveness of U-Print on app classification.

\begin{figure}
    \centering
    \includegraphics[width=1\linewidth]{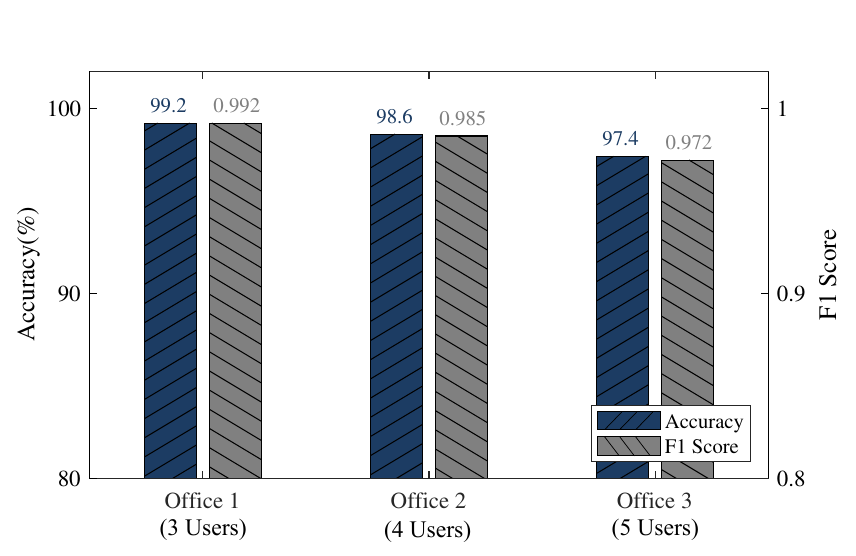}
    \caption{Identification performance in different environments.}
    \label{fig:user}
\end{figure}

\textbf{Performance of User Identification.} 
Then, we elaborated on U-Print's performance on user identification. 
Specifically, our system relies on behavior samples, consisting of app and action predictions from wireless traffic, to profile users' smartphone usage patterns.
Hence, we evaluated U-Print on Dataset C, a user profiling and identification dataset collected from three different office rooms, involving 12 users.
The system's performance in terms of users and environments is reported in Fig.~\ref{fig:allusers} and Fig.~\ref{fig:user}, respectively.

Fig.~\ref{fig:allusers} shows that U-Print achieved an identification accuracy of 95\% for most of the users.
Moreover, it can be observed that the identification accuracy of user 9 and user 11 was relatively low.
We suspected that the two users had similar add-on apps on their smartphones and exhibited analogous usage patterns, making our system harder to discriminate between them.
However, the overall user identification performance of our system still achieved 98\%.
Fig.~\ref{fig:user} depicts the performance of user identification in three environments.
We observed that the identification accuracy in each scenario had reached a level exceeding 97\%.
The average accuracy for user identification in Office 1 and Office 2 was even over 98.9\%, with an F1 score of approximately 0.99. 
In addition, U-Print had the lowest performance in Office 3, because there were five users in this office.
Despite that, our system still had a high accuracy of 97.4\% and an F1 score of 0.972.
The above results demonstrate the high effectiveness of U-Print in user identification based on wireless traffic. 
Moreover, ambient IoT traffic would have a limited influence on our system.
Generally, IoT traffic often exhibits distinguishable periodic patterns, such as fixed-interval pings or status updates, thus differing from human-driven app interactions~\cite{majumdar2020real}.
In this way, most of them would be classified as unknown by our app and action classifiers and clustered into meaningless user profiles by the K-Means algorithm.
Such profiles can be easily recognized and removed by attackers.

\begin{figure}
    \centering
    \includegraphics[width=1\linewidth]{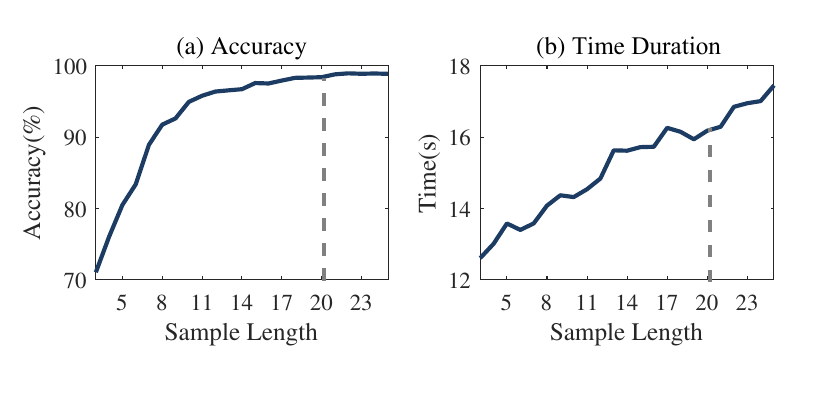}
    \caption{Identification accuracy and average time duration under different behavior sample lengths.}
    \label{fig:behavior}
\end{figure}

\textbf{Impact of Behavior Sample Length.} 
It is critical to determine the shortest sample length that enables accurate user identification.
Generally, longer behavior samples contain more information about users' smartphone usage habits.
However, excessively long samples may introduce redundant information and increase time and computational complexity.
To find a suitable sample length, we varied $W_b$ from 3 to 25 and recorded U-Print's identification accuracy and average time duration per sample in each setting.
The experimental results are shown in Fig.~\ref{fig:behavior}.
We observed that both the identification accuracy and time duration increased as the sample length became larger. 
From Fig.~\ref{fig:behavior}~(a), we found that the identification accuracy improved quickly when the sample length increased from 3 to 20.
However, when it went beyond 20, the accuracy stayed at the same level around 98\%.
In addition, Fig.~\ref{fig:behavior}~(b) indicates that the time duration has an approximately linear relationship with the sample length.
Considering the above observations, we set the sample length $W_b = 20$ to achieve a balance between system effectiveness and efficiency

\begin{figure}[t]
  \centering
  \includegraphics[width=1\linewidth]{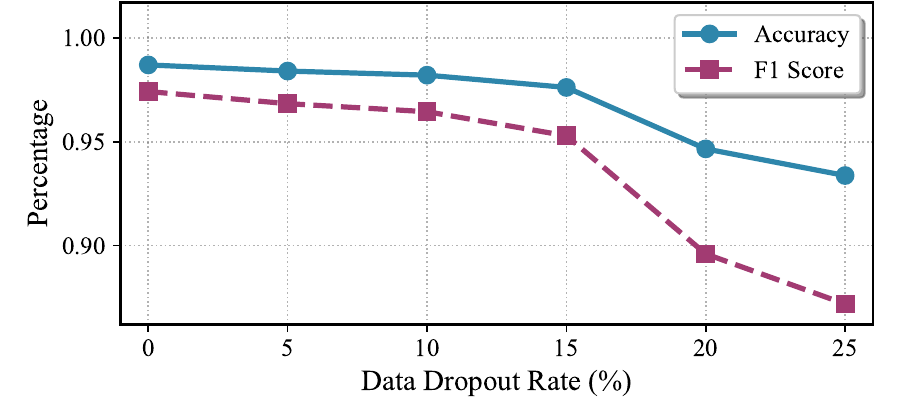} % myplot.eps
  \caption{Impact of data dropout.}
  \label{fig:datadrop_performance_analysis}
\end{figure}

\textbf{Impact of Packet Loss.}
Generally, the lower the data fidelity is, the more packets are lost.
We measured the data fidelity of the collected traffic using the 802.11 sequence number field and obtained an average fidelity of 65.83\%, which corresponds to a frame loss rate of 34.17\%.
Under this condition, our app classifier achieved an accuracy of over 98\% based on 30 received packets, and our action classifier obtained an accuracy of over 96\% based on traffic segments lasting 5 seconds each, demonstrating the robustness of our system under imperfect data capture conditions and its tolerance to real-world levels of packet loss.

To further study the impact of packet loss, we conduct an additional experiment by randomly dropping different percentages of collected packets from Dataset A.
As shown in Fig.~\ref{fig:datadrop_performance_analysis}, U-Print maintained strong performance under moderate packet loss.
Its accuracy and F1 score remained above 0.95 for dropout rates up to 15\%. 
The system performance began to degrade noticeably only when the dropout rate exceeded 20\%, with an accuracy of 94\% and an F1 score of 0.87 at a dropout rate of 25\%.
These findings confirm that U-Print's reliance on aggregated MAC-layer traffic patterns and temporal features makes it robust to occasional missing frames.

\begin{figure}[t]
  \centering
  \includegraphics[width=1\linewidth]{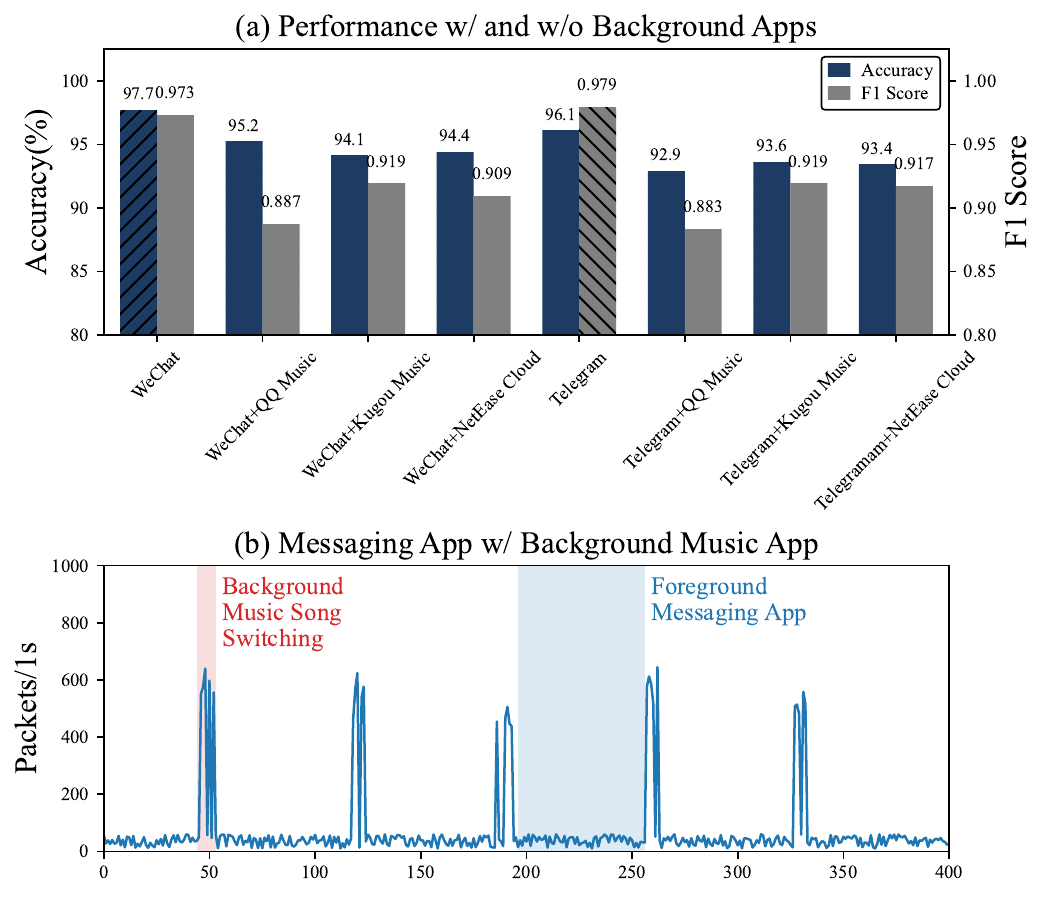} % myplot.eps
  \caption{Impact of background apps.}
  \label{fig:bg_traffic}
\end{figure}

\textbf{Impact of Background Apps.}
We evaluated the performance of U-Print in recognizing two messaging apps, namely WeChat and Telegram, when three music apps, i.e., QQ Music, Kugou Music, and NetEase Cloud, were running in the background.
This setting is representative and closely resembles real-world scenarios, and other app usage combinations are less likely to occur in practice.
For example, most video apps would disable the background music app automatically.
As shown in Fig.~\ref{fig:bg_traffic}~(a), with a background music app, an average accuracy reduction of approximately 3.1\% was shown in recognizing WeChat, and that of 2.8\% in predicting Telegram. 
Despite that, U-Print's performance remained within roughly 92\%-95\% in accuracy and 0.88-0.92 in F1 score. 
Moreover, we illustrated a trace of mixed wireless traffic in Fig.~\ref{fig:bg_traffic}~(b).
Generally, the packet rate of messaging apps was flat with a low variance.
However, some high-amplitude traffic bursts were appearing periodically.
These transient bursts are caused by song-switching events with buffering and asset fetching in the background.
But once the bursts subsided, the variations of the collected traffic returned to the normal level.
Thus, background traffic would bring a negative yet limited impact on our system.

\begin{figure}[t]
  \centering
  \includegraphics[width=1\linewidth]{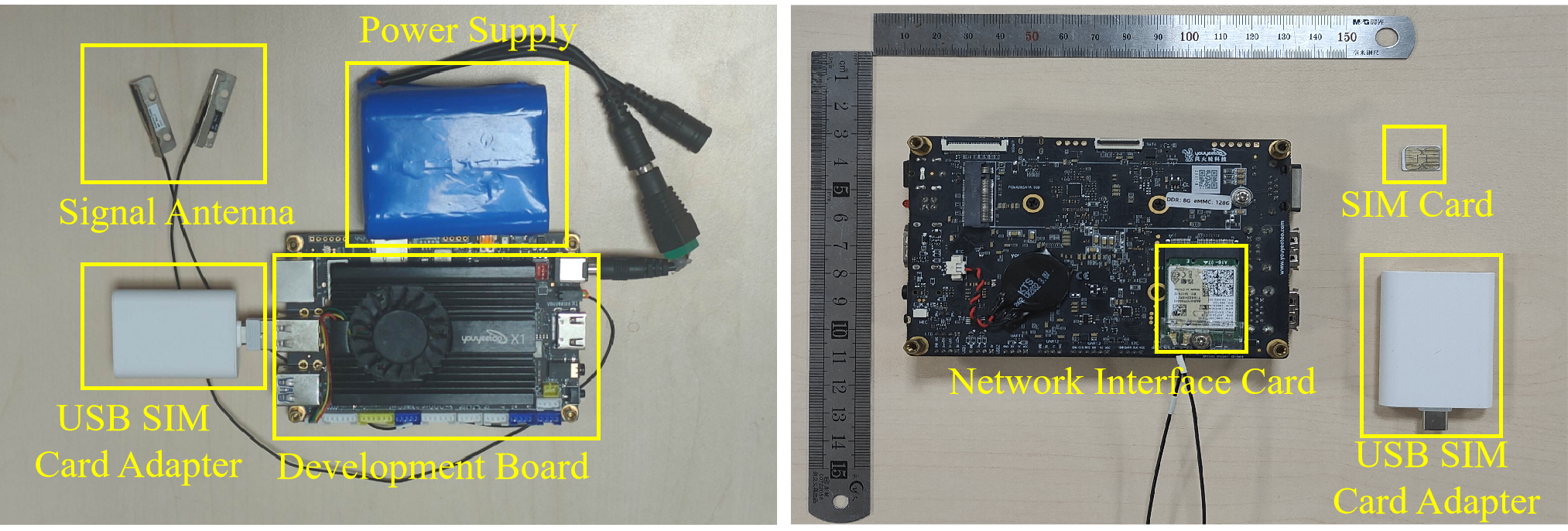} % myplot.eps
  \caption{Implementation of U-Print on a lightweight platform.}
  \label{fig:lightweightplatform}
\end{figure}

\textbf{Lightweight Implementation.}
To illustrate the concealability of U-Print, we built a lightweight prototype using a YOUYEETOO X1 development board, an Intel WiFi 6E AX201 NIC, a USB SIM card adapter, and a 12V power supply, as shown in Fig.~\ref{fig:lightweightplatform}. 
The development board had an 8G RAM and a size of 11.5 $\times$ 7.5 cm and ran Ubuntu 22.04 LTS.
The USB SIM card adapter was connected to the development board via a USB Type-C port and enabled LTE connectivity.
In this configuration, U-Print captured traffic and performed feature extraction, app and action classification, and user profiling and identification on the board.
Then, it forwarded the system results to a remote device. 
This approach reduces the on-site computational and size requirements and enables a more portable and concealable setup.

\begin{table}[t]
\centering
\caption{Runtime of U-Print on Different Platforms.}
\label{tab:runtime_comparison}
\resizebox{\columnwidth}{!}{%
\begin{tabular}{lcc}
\hline
\textbf{Component} & \textbf{Laptop} & \textbf{Development Board} \\
\hline
Traffic Preprocessing \& Feature Extraction & 0.042 ms & 0.099 ms \\
App \& Action Recognition  & 0.012 ms & 0.046 ms \\
User Profiling \& Identification & 0.125 ms & 0.331 ms \\
\hline
Total & 0.179 ms & 0.476 ms \\
\hline
\end{tabular}
}
\end{table}

\textbf{Time Consumption.}
To verify the real-time ability of U-Print, we demonstrated its runtime on the Lenovo laptop and the development board.
After implementing our system on the two platforms, we fed 1K samples into our system, and recorded the average runtime of three components, i.e., traffic preprocessing and feature extraction, app and action recognition, and user profiling and identification, for each test sample.
As reported in Table~\ref{tab:runtime_comparison}, the laptop was faster than the development board in each component, because it had better computational ability.
Moreover, most of the time consumption was attributed to user profiling and identification, about 0.125~ms on the Lenovo laptop and 0.331~ms on the development board.
The total time on the laptop and the development board was 0.179~ms and 0.476~ms per sample, respectively.
The above results suggest that our system has a quick response for each sample.

\section{Related Work}

\textbf{User Privacy Attacks on the MAC Layer.} 
With the proliferation of wireless devices, growing attention has been devoted to user privacy attacks on the MAC layer.
Zhang et al.~\cite{zhang2011inferring} proposed a hierarchical classification model that exploits statistical characteristics of Wi-Fi frames to recognize user online actions, such as web browsing, online gaming, video watching, etc. 
Moreover, Wang et al.~\cite{wang2015know} leveraged a random forest model to infer smartphone apps from wireless traffic.
Atkinson et al.~\cite{atkinson2018your} first recognized types of mobile apps and thus inferred potential user privacy information like age, gender, religion, etc.
However, the above approaches focused on app classification and were based on the closed-world assumption.
To transcend these limitations, Li et al.~\cite{li2022packet} recognized mobile apps of interest and in-app actions in the open-world setting using the sequential XGBoost and hierarchical bag-of-words models. 
However, all existing user privacy attacks on the MAC layer are limited to app and action recognition and cannot perform user identification.
On the contrary, U-Print not only recognized mobile apps and in-app actions in the open-world scenario but also identified smartphone users based on users' smartphone usage patterns.

\textbf{User Privacy Attacks on the Upper Layers.} 
Apart from the MAC layer, many researchers have paid attention to user privacy attacks on the upper layers.
Taylor et al.~\cite{taylor2016appscanner} built app fingerprints and recognized Android apps in real time, even though network traffic adopts payload encryption protocols such as HTTPS/TLS.
Xiang et al.~\cite{xiang2018appclassifier} used heuristic-based methods to re-correct mislabeled TCP traffic flow to realize real-time traffic inference under background noise interference.
Van Ede et al.~\cite{van2020flowprint} generated app fingerprints and recognized apps by automatically finding temporal correlations among destination-related features from TCP/UDP network flows.
Li et al.~\cite{li2022foap} used structural similarity to filter out traffic segments unrelated to the app of interest for improving app classification in the open world. 
However, these approaches require a direct connection to wireless APs or switches, which is hard to achieve due to physical isolation and other protective measures on network devices.

\section{Discussion}
\textbf{Limitations.} 
Due to the heavy background traffic or user multitasking, overlapped traffic traces are hard for our system to recognize.
Because wireless packets cannot be associated with specific flows or apps based on MAC frame fields, traffic patterns of different apps and in-app actions are confused.
Especially when background traffic is heavy, the traffic features of foreground apps are overwhelmed.
However, heavily overlapped traffic generally accounts for a small proportion in practice.
Additionally, app version updates, such as user interface (UI) or user experience (UX) changes, could cause protocol or data structure changes and thus impact system performance.
However, as long as the core communication module remains the same, such changes will not alter underlying traffic patterns.
In this case, our app and action classifiers remain effective in recent app versions.
For long-term adaptation, the attacker can periodically retrain our classifiers.

\textbf{Countermeasures.}
To defend against WiFi MAC-layer fingerprinting attacks like U-Print, dummy packet injection, frame padding and fragmentation, and channel hopping are possible solutions.
Dummy packet injection works by inserting fake yet legitimate frames into wireless traffic, thus disrupting communication patterns.
However, this approach incurs additional bandwidth and energy costs and may be detected by anomaly-based systems.
Frame padding and fragmentation obscure size-based features by transforming all packets into new packets with a fixed length before transmission, thereby undermining the size distributions.
Yet, this method introduces additional latency and processing overhead. 
Channel hopping disperses communication across multiple frequencies. 
This approach prevents an attacker from capturing complete traffic traces when sniffing a user, but it requires tight synchronization between the sender and receiver and reduces throughput due to increased channel contention.

\section{Conclusion}
This paper presents U-Print, a novel user privacy attack system that can passively construct user profiles and identify smartphone users by analyzing encrypted traffic at the MAC layer.
We observe that mobile users are tempted to install different apps based on their interests and exhibit unique usage patterns in the same app.
Our system can extract fine-grained features of wireless traffic, recognize add-on apps and in-app actions in the open world, and thus construct user fingerprints. 
We implement U-Print and evaluate it in real-world environments. 
The evaluation results demonstrate that U-Print can achieve a user identification accuracy of 98.4\%. 
In addition, U-Print can efficiently recognize apps and actions performed by smartphone users. 
This paper proposes and proves a reasonable and obvious threat to smartphone user privacy, and we hope that this study can enhance public awareness of privacy protection and promote new defenses against such attacks.

In future work, we aim to enhance system robustness under user multitasking scenarios by exploring fine-grained traffic separation techniques. 
We also plan to investigate the periodic updating scheme of our system against app version updates, as well as the defense mechanisms of user fingerprinting.

\printbibliography
% \bibliographystyle{IEEEtran}
% \bibliography{ref}
\end{document}